\documentclass[a4paper,11pt]{article}
\pdfoutput=1 

\usepackage{jcappub} 

\usepackage[T1]{fontenc} 

\title{\boldmath Towards an accurate treatment of the reduced speed of light approximation in parameterized radiative transfer simulations of reionization}

\author[a]{Christopher Cain\note{Corresponding author.}}

\affiliation[a]{School of Earth and Space exploration, Arizona State University, Tempe, AZ 85281, USA}

\emailAdd{clcain3@asu.edu}

\abstract{

The reduced speed of light approximation (RSLA) has been employed to speed up radiative transfer simulations of reionization by a factor of $\gtrsim 5-10$.  However, it has been shown to cause significant errors in the HI-ionizing background near reionization's end in simulations of representative cosmological volumes.  We show that using the RSLA is, to a good approximation, equivalent to re-scaling the global ionizing emissivity in a redshift-dependent way.  We derive this re-scaling and show that it can be used to ``correct'' the emissivity in RSLA simulations.  This method requires the emissivity to be re-scaled after the simulation has been run, which limits its applicability to situations where the emissivity is set ``by hand'' or determined by free parameters.  We test our method by running full speed of light simulations using these re-scaled emissivities and comparing them with their RSLA counterparts.   We find that for reduced speeds of light $\tilde{c} \geq 0.2$, the 21 cm power spectrum at $0.1 \leq k /[h{\rm Mpc}^{-1}] \leq 0.2$ and key Ly$\alpha$ forest observables agree to within $20\%$, and often within $10\%$, throughout reionization.  Position-dependent time-delay effects cause inaccuracies in reionization's morphology on large scales at the factor of $2$ level for $\tilde{c} \leq 0.1$.  Our method allows for up to a factor of $5$ speedup in studies that express the emissivity in terms of free parameters, including efforts to constrain the emissivity using observations.  This is a crucial step towards constraining the ionizing properties of high-redshift galaxies using efficient radiative transfer simulations.  }

\begin{document}
\maketitle
\flushbottom

\section{Introduction}
\label{sec:intro}

Radiative transfer (RT) simulations are a crucial tool for studying the Epoch of Reionization (EoR).  During this period, the first generation of hydrogen ionizing sources drove ionization fronts (I-fronts) into the neutral intergalactic medium (IGM), which later overlapped and produced a fully ionized IGM by $z = 5-6$~\citep{Fan2006,Robertson2015,Kulkarni2019}.  RT simulations can self-consistently track the growth and overlap of ionized regions during reionization, and model in detail the physical conditions in the intergalactic medium (IGM) during the process~\citep{Gnedin2014,Ocvirk2018,Keating2019,Garaldi2022}.  They also allow the physics of galaxies to be directly linked to IGM conditions, enabling observables that probe reionization to constrain the physics of high-redshift galaxies.  

Despite being the most physically accurate method available to study reionization, RT simulations are also the most computationally expensive.  Formally, the RT equation is a $7$-dimensional problem - three spatial, two angular, one frequency, and one time.  Moreover, representative cosmological volumes ($\gtrsim [100 {\rm Mpc}]^3$) typically contain millions of ionizing sources, further increasing the cost of RT.  Several methods, including moment-based RT~\citep{Gnedin2001,Aubert2008,Wu2021} and adaptive ray tracing~\citep{Abel2002,Trac2007} have been developed to speed up the angular component of the problem and eliminate the direct scaling of computation time with the number of sources.  For many applications, only one to a few frequency bins are required to accurately model IGM conditions~\citep[e.g.][]{Ocvirk2016,Keating2019,Asthana2024}.  Sub-grid treatments of the clumping of the IGM on $\sim$ kpc scales~\citep[e.g.][]{Mao2019,Cain2021} can allow for simulations to be run at relatively coarse spatial resolution, which helps to counter the computational scaling with number of spatial dimensions.  

An common trick used to speed up the time-dependence of the RT equation is the reduced speed of light approximation (RSLA).  The time step required to resolve the transfer of radiation on an Eulerian grid is 
\begin{equation}
    \label{eq:t_rad}
    t_{\rm rad} \leq \frac{\Delta x_{\rm cell}}{c}
\end{equation}
where $\Delta x_{\rm cell}$ is the cell width and $c$ is the speed of light.  Often, $t_{\rm rad}$ is much smaller than any other timescale (such as hydro-dynamical or chemical timescales), meaning that it could be much larger without affecting the treatment of these other processes.  Indeed, if the RT is done in post-processing (as in many applications), the only relevant timescales are those associated with the RT equation itself.  In addition to $t_{\rm rad}$, these are (1) the cell-crossing time of ionization fronts (I-fronts) and (2) the timescale over which the source population and underlying density field evolve.  The typical speed of I-fronts ($v_{\rm IF}$) is $1-2$ orders of magnitude lower than $c$~\citep{DAloisio2019}, and the galaxy population evolves on timescales of $10s-100s$ of Myr, much longer than typical values of $t_{\rm rad}$ ($\lesssim 1$ Myr).  

The RSLA introduces a smaller speed of light, $\tilde{c} < c$, chosen as small as possible so that $t_{\rm rad}$ remains the smallest timescale in the problem.  Because the number of time-steps is inversely proportional to $t_{\rm rad}$, this gives a direct computational speedup by a factor of $c/\tilde{c}$.  The RSLA has been used in a number of contexts to dramatically increase computational efficiency of reionization RT simulations at various spatial scales~\citep{DAloisio2020,Kannan2022}.  

However, several recent works~\citep{Gnedin2016,Deparis2019,Ocvirk2019,Cain2023} have noted that the RSLA fails in certain important regimes of reionization modeling.  Ref.~\cite{Gnedin2016} argued that the RSLA should be used only to model local fluctuations in the ionizing background during reionization, but never the mean ionizing background itself.  Supporting this point, Ref.~\cite{Ocvirk2019} found that using the RSLA leads to an under-estimate of the ionizing background close to and after the end of reionization by as much as a factor of $\tilde{c}/c$.  Most recently, Ref.~\cite{Cain2023} found that reionization models calibrated to reproduce recent Ly$\alpha$ forest observations at $5 < z < 6$(from Ref.~\citep{Bosman2021}) using the RSLA yielded biased results for the galaxy ionizing emissivity required to match these measurements.  In this work, we seek to understand the reasons for these issues and find a way to account for them.  

The findings of Ref.~\cite{Cain2023} hint at a possible pathway to a solution.  They found that in simulations with different values of $\tilde{c}$, the global ionizing emissivity $\dot{N}_{\rm em}(z)$ could be calibrated to reproduce measurements of the Ly$\alpha$ forest transmission at $z \leq 6$ (their Fig. 10).  This suggests that perhaps the RSLA is, at least approximately, equivalent to a redshift-dependent re-scaling of $\dot{N}_{\rm em}$.  In this work, we will show that this is indeed the case, and we will provide a formula for this re-scaling that can be applied to $\dot{N}_{\rm em}$ for any simulation that uses the RSLA.  In situations where $\dot{N}_{\rm em}$ is being set by hand (as in Ref.~\cite{Cain2023}), or is a function of some set of free parameters, this re-scaling can be applied to $\dot{N}_{\rm em}$ in ``post-processing''.  Then, the simulation can be {\it treated as if} it had used the full speed of light and the re-scaled $\dot{N}_{\rm em}$.  

This work is organized as follows: in \S\ref{sec:method}, we use a simple analytic model to understand why the RSLA fails in certain situations, and derive the aforementioned re-scaling of $\dot{N}_{\rm em}$.  In \S\ref{sec:numerical_methods}, we briefly describe our numerical methods for running RT simulations of reionization.  In \S\ref{sec:testing_method}-\ref{sec:multi-freq} we use simulations to test how well our approach works, and we conclude in \S\ref{sec:conc}.   Throughout, we assume the following cosmological parameters: $\Omega_m = 0.305$, $\Omega_{\Lambda} = 1 - \Omega_m$, $\Omega_b = 0.048$, $h = 0.68$, $n_s = 0.9667$ and $\sigma_8 = 0.82$, consistent with Ref.~\cite{Planck2018} results. All distances are quoted in co-moving units unless otherwise specified.  Throughout, we will work in units in which the speed of light is unity ($c = 1$).  

\section{The Method}
\label{sec:method}

\subsection{When and why does the RSLA fail?}
\label{subsec:fail}

We will first build intuition for when and why the RSLA fails using a simple analytic argument.   In the early stages of reionization, the universe contains a collection of ionized ``bubbles'' surrounding isolated clusters of ionizing sources, and is otherwise fully neutral.  The mean photo-ionization rate, $\Gamma_{\rm HI}$, inside one of these bubbles is
\begin{equation}
    \label{eq:gammaHI}
    \Gamma_{\rm HI} = N_{\gamma} c \sigma_{\rm HI}
\end{equation}
where $N_{\gamma}$ is the mean number density of ionizing photons inside the bubble, $c$ is the speed of light, and $\sigma_{\rm HI}$ is the HI-ionizing cross-section averaged over the ionizing spectrum.  At some time $t$ after reionization starts, $N_{\gamma}$ is given by
\begin{equation}
    \label{eq:Ngamma}
    N_{\gamma}(t) = \int_{0}^t dt' [\dot{N}_{\rm em}(t') - \dot{N}_{\rm abs}(t')]
\end{equation}
where $\dot{N}_{\rm em}$ is the emission rate of photons inside the bubble by sources and $\dot{N}_{\rm abs}$ is the absorption rate.  The latter can be written as two terms, 
\begin{equation}
    \label{eq:Ndot_abs}
    \dot{N}_{\rm abs} = \dot{N}_{\rm abs}^{\rm ion} + \dot{N}_{\rm abs}^{\rm rec}
\end{equation}
where $\dot{N}_{\rm abs}^{\rm ion}$ is the absorption rate by neutral gas at the edges of the bubble, and $\dot{N}_{\rm abs}^{\rm rec}$ is due to recombinations within the bubble.  

We will first consider the limit where the absorption rate is dominated by ionizations of neutral H atoms, such that $\dot{N}_{\rm abs}^{\rm ion} >> \dot{N}_{\rm abs}^{\rm rec}$.  This limit is a good approximation early in reionization when the ionized bubbles are much smaller than their recombination-limited Stromgren volumes and most ionized regions have not yet overlapped.  The absorption rate can then be written as
\begin{equation}
    \label{eq:Ndot_abs_2}
    \dot{N}_{\rm abs}(t) = \dot{N}_{\rm abs}^{\rm ion}(t) \approx \dot{N}_{\rm em}(t - R/c)
\end{equation}
where we have approximated the ionized region as a spherical bubble with radius $R$.  The quantity $t - R/c$ is the so-called ``retarded time'' at the boundary of the ionized region (the ionization front or I-front) relative to the sources near the bubble's center.  That is, photons emitted at time  $t - R/c$ will reach the bubble's edge and be absorbed at time $t$.   

We next assume that $R/c$ is sufficiently small that both $\dot{N}_{\rm em}$ and $R$ can be approximated as constant across this length of time.  In this limit, Eq.~\ref{eq:Ngamma} evaluates to
\begin{equation}
    \label{eq:Ngamma_2}
    N_{\gamma} = \dot{N}_{\rm em} \times (t - [t - R/c])  = \dot{N}_{\rm em} \frac{R}{c}
\end{equation}
Putting this into Eq.~\ref{eq:gammaHI}, we get
\begin{equation}
    \label{eq:gammaHI_local_source_approx}
    \Gamma_{\rm HI} = \dot{N}_{\rm em} \frac{R}{c} c \sigma_{\rm HI} = \dot{N}_{\rm em} R \sigma_{\rm HI}
\end{equation}
In this limit, $\Gamma_{\rm HI}$ is independent of the speed of light, and we can get the right $\Gamma_{\rm HI}$ inside the bubble using the RSLA.  Indeed, replacing $R$ in Eq.~\ref{eq:gammaHI_local_source_approx} with the mean free path to ionizing photons, $\lambda$, reveals that we have reproduced the so-called ``local source approximation'' with its well-known relationship between $\Gamma_{\rm HI}$, $\lambda$, and $\dot{N}_{\rm em}$~\citep[e.g. Ref.][]{Becker2013}.  

Next, we consider the opposite limit, in which $\dot{N}_{\rm abs}^{\rm ion} = 0$.   In this case, either the bubble has reached its limiting Stromgren volume (in which case $\dot{N}_{\rm abs}^{\rm rec} \approx \dot{N}_{\rm em}$), or the universe is entirely ionized.  Either way, the right-hand side of Eq.~\ref{eq:Ngamma} will be independent of the speed of light, since neither $\dot{N}_{\rm em}$ nor $\dot{N}_{\rm abs}^{\rm rec}$ depend on it\footnote{In principle, the recombination rate could be affected by $c$ if self-shielding effects are at play, since a difference in $\Gamma_{\rm HI}$ can change the self-shielding properties of the IGM~\citep{DAloisio2020,Nasir2021,Theuns2023}.  However, we expect this to be a small effect in the context of this work, since we will be comparing simulations that have fairly similar mean $\Gamma_{\rm HI}$.  }.  Thus, $N_{\gamma}$ is independent of $c$, and we have 
\begin{equation}
    \label{eq:gammaHI_rec_only_approx}
    \Gamma_{\rm HI} \propto c
\end{equation}
Thus, using the RSLA in this limit would result in $\Gamma_{\rm HI}$ being incorrect by the same factor by which $c$ is reduced.  

The first approximation - that $\dot{N}_{\rm abs}^{\rm ion}$ dominates the absorption rate - is valid near the beginning of reionization, as we mentioned earlier.  The reason $\Gamma_{\rm HI}$ is independent of $c$ in this limit is that the time required for photons to reach neutral gas is proportional to $1/c$, such that $N_{\gamma}$ inside the bubble(s) is also proportional to this factor. This cancels the factor of $c$ in Eq.~\ref{eq:gammaHI}, which gives the correct $\Gamma_{\rm HI}$ if a reduced $c$ is used, even though $N_{\gamma}$ is incorrect.  In the other limit, the fact that $N_{\gamma}$ is independent of $c$ means that it will still be correct when a reduced $c$ is used, causing $\Gamma_{\rm HI}$ to be incorrect.  The first limit is expected to apply near the beginning of reionization, and the second near and after its end.  The limiting behavior described here is exactly what was found in Ref.~\cite{Ocvirk2019} (see also Ref.~\cite{Deparis2019}).  

\subsection{Fixing the problem}
\label{subsec:fix}

In this section, we will show that, to a good approximation, the use of the RSLA equivalent to a redshift-dependent re-scaling of $\dot{N}_{\rm em}$.  Consider two simulations, one using the full speed of light $\tilde{c} = c = 1$ and the other a reduced speed of light $\tilde{c} < 1$.  Let these simulations have global ionizing emissivities $\dot{N}_{\rm em}^{c}$ and $\dot{N}_{\rm em}^{\tilde{c}}$, respectively.  Our goal is to find a relationship between these four quantities such that these two simulations have identical physical properties - including reionization history, net absorption rate, $\Gamma_{\rm HI}$, etc.  In particular, if they have the same absorption rate (that is, $\dot{N}_{\rm abs}^{c} = \dot{N}_{\rm abs}^{\tilde{c}}$), then from Eq.~\ref{eq:Ngamma}, we have
\begin{equation}
    \label{eq:Nabs_dot_eq}
    N_{\gamma}^{c}(t) - \int_{0}^t dt' \dot{N}_{\rm em}^{c}(t') = N_{\gamma}^{\tilde{c}}(t) - \int_{0}^t dt' \dot{N}_{\rm em}^{\tilde{c}}(t')
\end{equation}
where we have used the same speed-of-light notation for $N_{\gamma}$ that we did for $\dot{N}_{\rm em}$.  If we also assume that both simulations have the same mean $\Gamma_{\rm HI}$, then from Eq.~\ref{eq:gammaHI} we have
\begin{equation}
    \label{eq:Ngamma_eq}
    N_{\gamma}^{c} c \sigma_{\rm HI} = N_{\gamma}^{\tilde{c}} \tilde{c} \sigma_{\rm HI} \rightarrow N_{\gamma}^{c} = N_{\gamma}^{\tilde{c}} \frac{\tilde{c}}{c}
\end{equation}
Provided Eq.~\ref{eq:Ngamma_eq} holds at all times, we can differentiate both sides with respect to time to get
\begin{equation}
    \label{eq:Ngamma_dot_eq}
    \dot{N}_{\gamma}^{c} = \dot{N}_{\gamma}^{\tilde{c}} \frac{\tilde{c}}{c}
\end{equation}
Differentiating both sides of Eq.~\ref{eq:Nabs_dot_eq} with respect to time and substituting for $N_{\gamma}^{c}$ using Eq.~\ref{eq:Ngamma_dot_eq} yields the main result of this paper, 
\begin{equation}
    \label{eq:main_result}
    \boxed{\dot{N}_{\rm em}^{c}(t) = \dot{N}_{\rm em}^{\tilde{c}}(t) - \dot{N}_{\gamma}^{\tilde{c}}(t) \left(1 - \frac{\tilde{c}}{c}\right)}
\end{equation}
We have explicitly written the time dependence to emphasize that Eq.~\ref{eq:main_result} holds at all times.  

Eq.~\ref{eq:main_result} expresses the emissivity in the simulation with the full speed of light in terms of $c$, $\tilde{c}$, $N_{\rm em}^{\tilde{c}}$, and $\dot{N}_{\gamma}^{\tilde{c}}$.  The first two of these are just numbers, and the other two are known after the simulation with the reduced speed of light has been run.  This allows us to calculate $\dot{N}_{\rm em}^{c}$ {\it without running any simulations with the full speed of light}.  Again, $\dot{N}_{\rm em}^{c}$ is the emissivity would produce (approximately) the same ionization history, $\Gamma_{\rm HI}$, etc. as the reduced speed of light run, had the full speed of light been used.  Of course, since we had to assume that this relationship exists in order to derive Eq.~\ref{eq:main_result}, we have not yet proven that this statement is true.  We will do so using RT simulations in \S\ref{sec:testing_method}-\ref{sec:multi-freq}.  

The usefulness of Eq.~\ref{eq:main_result} is found in situations where $\dot{N}_{\rm em}$ is being treated as a ``free parameter'', which can be ``corrected'' after the simulation has been run.  Indeed, in works such as Refs.~\cite{Kulkarni2019,Keating2019,Cain2021}, $\dot{N}_{\rm em}$ is a ``free function'' of redshift that is set by hand to match some set of observables - in those works, the mean transmission of Ly$\alpha$ forest at $z \leq 6$. In many semi-numerical frameworks~\citep[e.g. 21cmFAST, Refs.][]{Mesinger2007,Qin2021}, $\dot{N}_{\rm em}$ is uniquely determined by a set of free parameters that are being constrained using a set of observations.  In this case, a re-scaling of $\dot{N}_{\rm em}$ corresponds to changing the mapping between these parameters and $\dot{N}_{\rm em}$ in a way that can be quantified.  We emphasize the Eq.~\ref{eq:main_result} {\it does not} constitute a correction to the RSLA simulation results themselves, but rather simply a re-scaling of the global emissivity after the simulation has been run.  As such, it is unfortunately not directly applicable in situations where $\dot{N}_{\rm em}$ is a self-consistent prediction of e.g. an underlying galaxy model~\cite{Ocvirk2018,Kannan2022,Garaldi2022}.  In this situation, one would require a method to correct the simulation quantities themselves, which is beyond the scope of this work.  
 
\section{Numerical Methods}
\label{sec:numerical_methods}

Our simulations are run with the adaptive ray-tracing RT code FlexRT, first described in~\citep{Cain2021,Cain2022b} and tested in Ref.~\cite{Cain2024}.  Here, we will briefly discuss the relevant features of the code, referring the reader to these works for details.  FlexRT solves the RT equation in post-processing on a time-series of cosmological density fields.  We model the ionizing sources in a simple manner by extracting halos from an N-body simulation with the same large-scale structure as the density field and assigning them ionizing emissivities (see below).  The RT equation itself is solved using an adaptive ray tracing method similar to the one described in Ref.~\cite{Abel2002} and employed in the code of Ref.~\cite{Trac2007}.  Sub-resolved ionization fronts are tracked using the ``moving-screen'' approximation, and the temperature behind them ($T_{\rm reion}$) is estimated using the flux-based method described in Ref.~\cite{DAloisio2019}\footnote{The subsequent temperature evolution in ionized cells is calculated using their Eq. 6.  }.  FlexRT solves for the opacity of the ionized IGM to ionizing photons using a novel sub-grid model based on high-resolution hydro/RT simulations of IGM gas dynamics similar to those described in~Refs.~\cite{DAloisio2020,Nasir2021}.  Recent improvements to this model, and details about forward-modeled observables are detailed in Ref.~\cite{Cain2023}.  

In \S\ref{sec:testing_method}, we will use the ``Reference'' setup in Ref.~\cite{Cain2023}, which assumes a fully dynamically relaxed IGM and mono-chromatic RT.  The physical properties of this model are summarized in their Fig. 2 and associated text.  Throughout in \S\ref{sec:testing_method}, the only parameters we will change relative to their Reference model are $\dot{N}_{\rm em}$ and $\tilde{c}$.  We will generalize and test Eq.~\ref{eq:main_result} in the context of multi-frequency RT simulations in \S\ref{sec:multi-freq}.  Our simulations have a box size of $L = 200$ $h^{-1}$Mpc with $N_{\rm RT} = 200^3$ RT cells.  Ly$\alpha$ forest properties are calculated (following \S3.7 of Ref.~\cite{Cain2023}) on a high-resolution ($N = 2048^3$) hydrodynamics simulation with the same large-scale initial conditions used for the N-body simulation from which we get the halos.  In this work, we will consider reduced speeds of light of $\tilde{c} = 0.05$, $0.1$, $0.2$, $0.3$, and $0.5$, alongside simulations that use the full speed of light ($\tilde{c} = c = 1$).

\section{Testing the Method}
\label{sec:testing_method}

\subsection{Strategy \& Terminology}
\label{subsec:terms}

The claim of Eq.~\ref{eq:main_result} can be summarized as follows: {\bf Given an RT simulation run with a reduced speed of light $\tilde{c}$ and emissivity $\dot{N}_{\rm em}^{\tilde{c}}(z)$ that has a mean photon number density $N_{\gamma}^{\tilde{c}}(z)$, a simulation using the full speed of light and emissivity $\dot{N}_{\rm em}^{c}(z)$ should have the same reionization history, ionization morphology, IGM transmission properties, etc. as the original simulation.}  We can verify whether this claim is true by comparing two types of simulations.  We first run simulations using the RSLA, with $\tilde{c} < 1$.  We refer to these simply as \textsc{reduced-$c$} runs, with emissivity $\dot{N}_{\rm em}^{\tilde{c}}$.  Next, we apply the re-scaling in Eq.~\ref{eq:main_result} to $\dot{N}_{\rm em}^{\tilde{c}}$ to obtain  $\dot{N}_{\rm em}^{c}$, and run another simulation with the full speed of light using $\dot{N}_{\rm em}^{c}$.  We refer to these as \textsc{full-$c$, re-scaled $\dot{N}_{\rm em}$} runs.  \textit{Eq.~\ref{eq:main_result} is accurate to the extent that the properties of these two types of simulations agree with each other.}  Lastly, for purposes of comparison, we will also include a third type of simulation, in which we use the  full speed of light but do not apply any re-scaling to $\dot{N}_{\rm em}^{\tilde{c}}$.  We refer to these as \textsc{full-$c$, un-scaled $\dot{N}_{\rm em}$} runs.  We summarize the properties of these three types of simulations in Table~\ref{tab:terminology}.  

\begin{table}
    \centering
    \begin{tabular}{|c||c|c|c|}
        \hline
        \bf Name & \bf Emissivity & \bf Speed of Light & \bf Description\\
        \hline
        \hline
         & & & Simulation using reduced \\ 
        \textsc{Reduced-$c$} & $\dot{N}_{\rm em}^{\tilde{c}}$ & $\tilde{c} < 1$  & speed of light using\\
         & & & un-scaled emissivity\\
        \hline
        & & & Full speed of light \\
        \textsc{full-$c$, re-scaled $\dot{N}_{\rm em}
        $}  & $\dot{N}_{\rm em}^{c}$ (Eq.~\ref{eq:main_result}) & $\tilde{c} = c = 1$  & simulation with $\dot{N}_{\rm em}$ \\ 
        & & & re-scaled using Eq.~\ref{eq:main_result}\\
        \hline
        & & &  Full speed of light\\
        & & & simulation, but using\\
        \textsc{full-$c$, un-scaled $\dot{N}_{\rm em}$}  &  $\dot{N}_{\rm em}^{\tilde{c}}$ & $\tilde{c} = c = 1$ & $\dot{N}_{\rm em}$ from the\\ 
        & & & \textsc{Reduced-$c$} run without\\ 
        & & & any re-scaling applied\\
         \hline
    \end{tabular}
    \caption{Summary of different kinds of simulations referred to in this work, and the names used for each (left-most column).  The second column from left gives the emissivity used in that simulation, and the third column the speed of light.  The right-most column briefly describes each type of simulation.  }
    \label{tab:terminology}
\end{table}

\subsection{Ionization history and morphology}
\label{subsec:ion_hist_morph}

We begin by studying how well the reionization history and morphology (shapes and sizes) of ionized regions match in the \textsc{reduced-$c$} and \textsc{full-$c$, re-scaled $\dot{N}_{\rm em}$} runs.  For this section, we scale up $\dot{N}_{\rm em}$ from the reference model of Ref.~\cite{Cain2023} until reionization ends at $z \approx 6$ when $c = 1$.  This is our \textsc{full-$c$, un-scaled $\dot{N}_{\rm em}$} run.  Next, we use the same $\dot{N}_{\rm em}$ to run a series of \textsc{reduced-$c$} runs with $\tilde{c} = 0.05$, $0.1$, $0.2$, $0.3$, and $0.5$.  Lastly, we evaluate Eq.~\ref{eq:main_result} for each of these and then run a set of \textsc{full-$c$, re-scaled $\dot{N}_{\rm em}$} simulations.  Again, the degree to which the \textsc{reduced-$c$} and corresponding \textsc{full-$c$, re-scaled $\dot{N}_{\rm em}$} runs agree with each other determines how accurate is Eq.~\ref{eq:main_result}.  

\begin{figure*}
    \centering
    \includegraphics[scale=0.145]{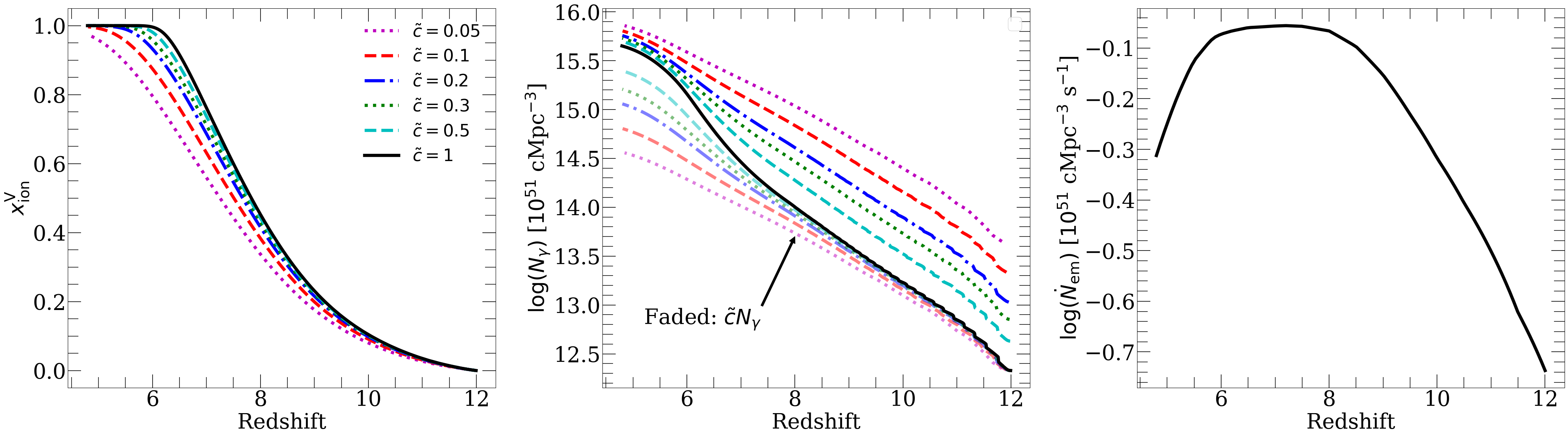}
    \caption{Summary of \textsc{full-$c$, un-scaled $\dot{N}_{\rm em}$} and \textsc{reduced-$c$} simulations used to study how well Eq.~\ref{eq:main_result} works with respect to the reionization history and morphology.  {\bf Left:} the reionization history for each value of $\tilde{c}$, using a scaled-up version of $\dot{N}_{\rm em}$ from the reference model from Ref.~\cite{Cain2023}, which ends reionization at $z = 6$ for $\tilde{c} = c = 1$ (black solid curve, the \textsc{full-$c$, un-scaled $\dot{N}_{\rm em}$} run).  Reducing $\tilde{c}$ pushes the end of reionization later, such that it has not finished by $z = 4.8$ for $\tilde{c} = 0.05$.  {\bf Middle:} number density of ionizing photons, $N_{\gamma}$.  The thick lines denote the actual values, while the faded lines show $\tilde{c} N_{\gamma}$.  Consistent with our analytic argument in \S\ref{subsec:fail}, at the beginning of reionization $\dot{N}_{\gamma}$ is a factor of $1/\tilde{c}$ higher in the \textsc{reduced-$c$} runs than in the \textsc{full-$c$, un-scaled $\dot{N}_{\rm em}$} case (Eq.~\ref{eq:Ngamma_2}).  Near reionization's end, they approach similar values.  {\bf Right:} Ionizing emissivity used in all these simulations ($\dot{N}_{\rm em}^{\tilde c}$).  }
    \label{fig:reion_models_1}
\end{figure*}

\subsubsection{Ionization history}
\label{subsubsec:ion_history}

In Figure~\ref{fig:reion_models_1}, we show the volume-averaged ionized fraction $x_{\rm ion}^{\rm V}$ (left), the ionizing photon number density $\dot{N}_{\gamma}$ (middle), and the ionizing emissivity $\dot{N}_{\rm em}$ (right) for our \textsc{full-$c$, un-scaled $\dot{N}_{\rm em}$} run (black solid curve) and our \textsc{reduced-$c$} runs (colored curves, see legend).  In the left panel, we see that the reionization histories are initially similar, but start to diverge as reionization enters its later stages.  The end of reionization is delayed by $\Delta z \approx 1.5$ in the $\tilde{c} = 0.05$ simulation relative to the $\tilde{c} = 1$ case.  The bold curves in the middle panel show $\dot{N}_{\gamma}$ for each simulation, while the faded curves show $\tilde{c} N_{\gamma}$.  Early in reionization, $\dot{N}_{\gamma}$ scales linearly with $1/\tilde{c}$, consistent with the early reionization limit described by Eq.~\ref{eq:Ngamma_2}.  However, as reionization ends, $\dot{N}_{\gamma}$ for all simulations approaches the same value, reflecting the behavior expected near reionization's end in \S\ref{subsec:fail}.  The right-most panel shows $\dot{N}_{\rm em}^{\tilde{c}}$, which is the same for all the simulations shown here.  

\begin{figure*}
    \centering
    \includegraphics[scale=0.21]{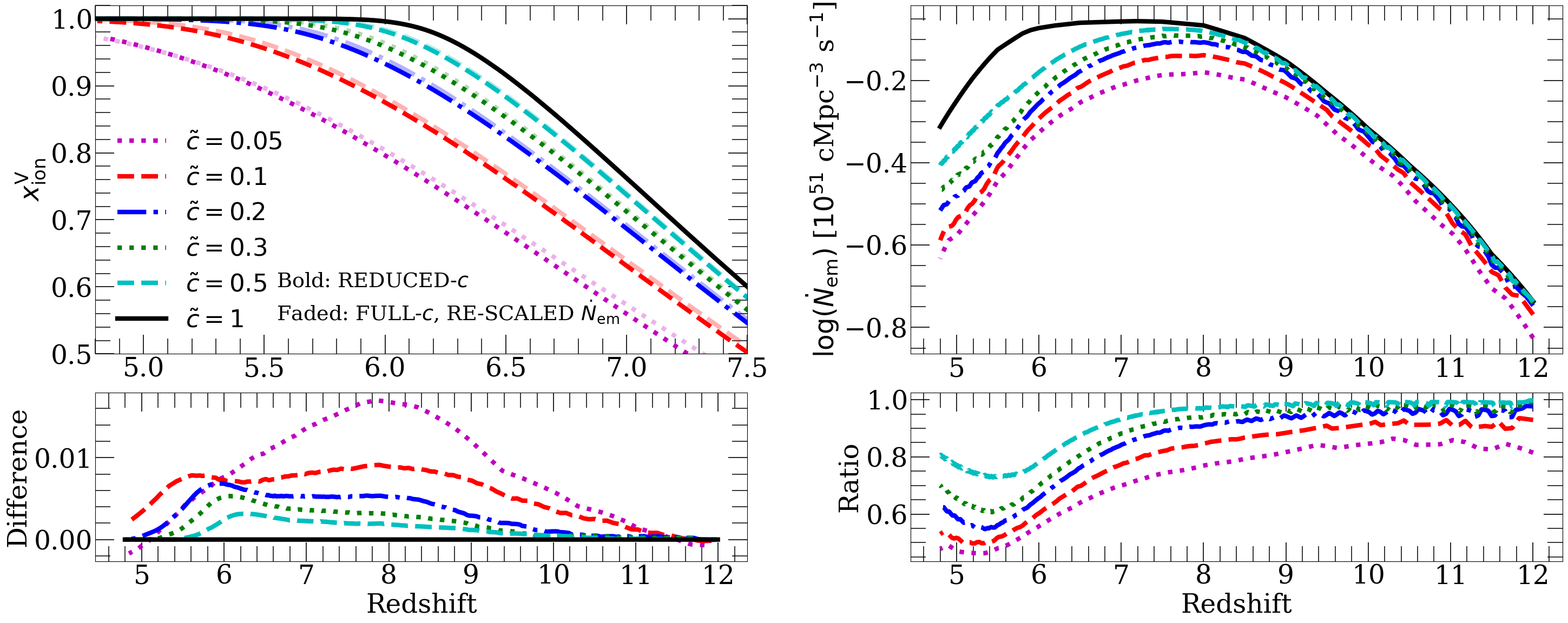}
    \caption{Comparison of the reionization history and emissivity in our \textsc{reduced-$c$} and \textsc{full-$c$, re-scaled $\dot{N}_{\rm em}$} simulations.  {\bf Upper Left:} the same reionization histories (again), but zoomed in on the last half of reionization and including the \textsc{full-$c$, re-scaled $\dot{N}_{\rm em}$} results for each value of $\tilde{c}$ (faded curves).  The bold and faded curves agree very well for all values of $\tilde{c}$, such that they are difficult to tell apart on the plot. {\bf Lower Left:} the linear differences between $x_{\rm ion}^{\rm V}$ for the \textsc{reduced-$c$} and corresponding \textsc{full-$c$, re-scaled $\dot{N}_{\rm em}$} runs.  This is less than $0.01$ for all except $\tilde{c} = 0.05$.  {\bf Upper Right:} emissivities for each of the \textsc{full-$c$, re-scaled $\dot{N}_{\rm em}$} runs ($\dot{N}_{\rm em}^{c}$ from Eq.~\ref{eq:main_result}, colored curves) compared to $\dot{N}_{\rm em}^{\tilde{c}}$ (black solid curve).  {\bf Bottom Right:} ratio of $\dot{N}_{\rm em}^{c}$ and $\dot{N}_{\rm em}^{\tilde{c}}$ for each $\tilde{c}$.  The ratio is smallest at the end of reionization, dropping to $0.5$ ($0.8$) for $\tilde{c} = 0.05$ ($0.5$).  }
    \label{fig:reion_models_2}
\end{figure*}

In the left panels Figure~\ref{fig:reion_models_2}, we compare each of our \textsc{reduced-$c$} runs to their \textsc{full-$c$, re-scaled $\dot{N}_{\rm em}$} counterparts.  We show the former as bold curves and the latter as faded curves, with the color and line style identifying the value of $\tilde{c}$.  The upper left panel shows $x_{\rm ion}^{\rm V}$, zoomed in on the second half of the reionization history for clarity.  We find excellent agreement between the two even for our lowest value of $\tilde{c} = 0.05$, such that the different sets of curves are difficult to see on the plot.  The bottom left panel shows the linear differences of $x_{\rm ion}^{\rm V}$ in the \textsc{reduced-$c$} and \textsc{full-$c$, re-scaled $\dot{N}_{\rm em}$} runs over the entire reionization history.  For all but the $\tilde{c} = 0.05$ case, the difference is always $\lesssim 0.01$, and for $\tilde{c} \geq 0.3$ it is $< 0.005$.  

In the upper right panel, we show $\dot{N}_{\rm em}^{c}$ calculated using Eq.~\ref{eq:main_result}, and the bottom right panel shows the ratio with $\dot{N}_{\rm em}^{\tilde{c}}$ (the black solid curve).  This ratio differs from unity the most at the end of reionization, dropping to $\approx 0.5$ ($0.8$) by $z = 5$ for $\tilde{c} = 0.05$ ($0.5$).  The fact that the \textsc{reduced-$c$} and \textsc{full-$c$, re-scaled $\dot{N}_{\rm em}$} runs have such different $\dot{N}_{\rm em}$ but similar reionization histories validates Eq.~\ref{eq:main_result}.  As we will see, the small differences in reionization history arise because Eq.~\ref{eq:main_result} applies to globally averaged quantities, and does not account for spatial fluctuations in the distribution of photons in the IGM.   

These results show that, at least with respect to the global ionization history, the claim of Eq.~\ref{eq:main_result} is reasonably accurate.  Namely, that using the RSLA is equivalent to a redshift-dependent re-scaling of $\dot{N}_{\rm em}$.   We emphasize again that Eq.~\ref{eq:main_result} does {\bf not} prescribe a way to directly correct the \textsc{reduced-$c$} simulation results for the effects of the RSLA.  Put another way, we do not prescribe a way to convert the \textsc{reduced-$c$} results into those of the \textsc{full-$c$, un-scaled $\dot{N}_{\rm em}$} results.  As such, our method is not directly applicable when $\dot{N}_{\rm em}$ is a {\it prediction} of e.g. some underlying galaxy model.  

\subsubsection{Ionization morphology}
\label{subsubsec:ion_morphology}

\begin{figure*}
    \centering
    \includegraphics[scale=0.132]{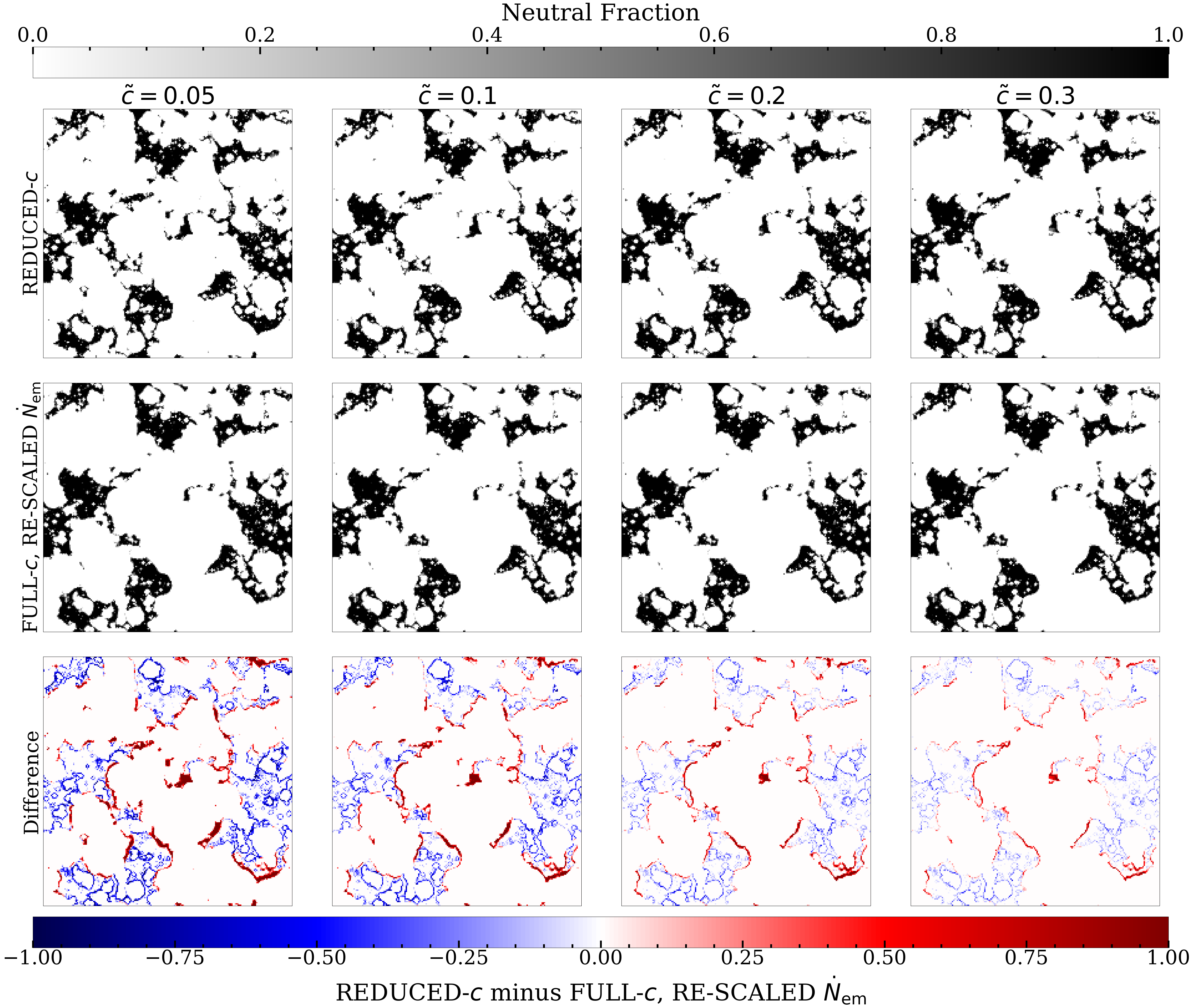}
    \caption{Visualization of the ionization field at $80\%$ ionized in \textsc{reduced-$c$} (top row) and \textsc{full-$c$, re-scaled $\dot{N}_{\rm em}$} runs (middle row) for $\tilde{c} = 0.05$, $0.1$, $0.2$, and $0.3$ (left to right).  We show the difference maps for each value of $\tilde{c}$ in the bottom row.  We see in the top row that as $\tilde{c}$ decreases (right to left), the neutral regions (islands) become more extended and porous at fixed ionized fraction.  We do not see this in the middle row, indicating that it is not due to differences in the reionization history between \textsc{reduced-$c$} runs.  Instead, it is because the RSLA induces a longer delay between emission and absorption for photons emitted in large vs. small ionized bubbles.  The re-scaling in Eq.~\ref{eq:main_result} corrects for the delay only on average, and does not account for position dependence.  The result is that the morphologies in the \textsc{reduced-$c$} and \textsc{full-$c$, re-scaled $\dot{N}_{\rm em}$} runs diverge as $\tilde{c}$ decreases.  }
    \label{fig:ion_field_vis}
\end{figure*}

Next, we apply the same analysis to the morphology of ionized and neutral regions, and statistics that depend on it.  Figure~\ref{fig:ion_field_vis} shows maps of the ionization field when the universe is $80\%$ ionized for \textsc{reduced-$c$} (top row) and \textsc{full-$c$, un-scaled $\dot{N}_{\rm em}$} runs (middle row).  Dark regions are neutral and white ones are ionized.  The bottom row shows the difference maps for the top and middle rows.  Red (blue) denotes regions that are more neutral (ionized) in the \textsc{reduced-$c$} simulations.  

From right to left (decreasing $\tilde{c}$) in the top row, the neutral regions (islands) grow slightly more extended and porous.  Equivalently, the smallest ionized bubbles grow larger, and the largest ones smaller, when $\tilde{c}$ decreases.  We see no such trend, however, in the middle row.  From this, we deduce that the morphological differences in the top row are not due to the \textsc{reduced-$c$} runs having different reionization histories, since such differences would appear for the \textsc{full-$c$, re-scaled $\dot{N}_{\rm em}$} also.  They are instead a position-dependent artifact of the RSLA.  Photons emitted in large ionized regions experience a longer delay between emission and absorption than do those emitted in small bubbles.  Eq.~\ref{eq:main_result} accounts for this time-delay only on average, and thus misses the differences between photons emitted in different regions.  As a result, the largest (smallest) bubbles are too small (large) in the \textsc{reduced-$c$} runs.  We can most clearly see this effect in the difference maps, where it is clearly visible even for $\tilde{c} = 0.3$.  In this work, we do not attempt to correct for this position-dependent effect, since doing so would be complicated by the non-locality of the ionizing radiation field.  As we will see, this effect will set the minimum value of $\tilde{c}$ for which Eq.~\ref{eq:main_result} can be used. 

\begin{figure*}
    \centering
    \includegraphics[scale=0.105]{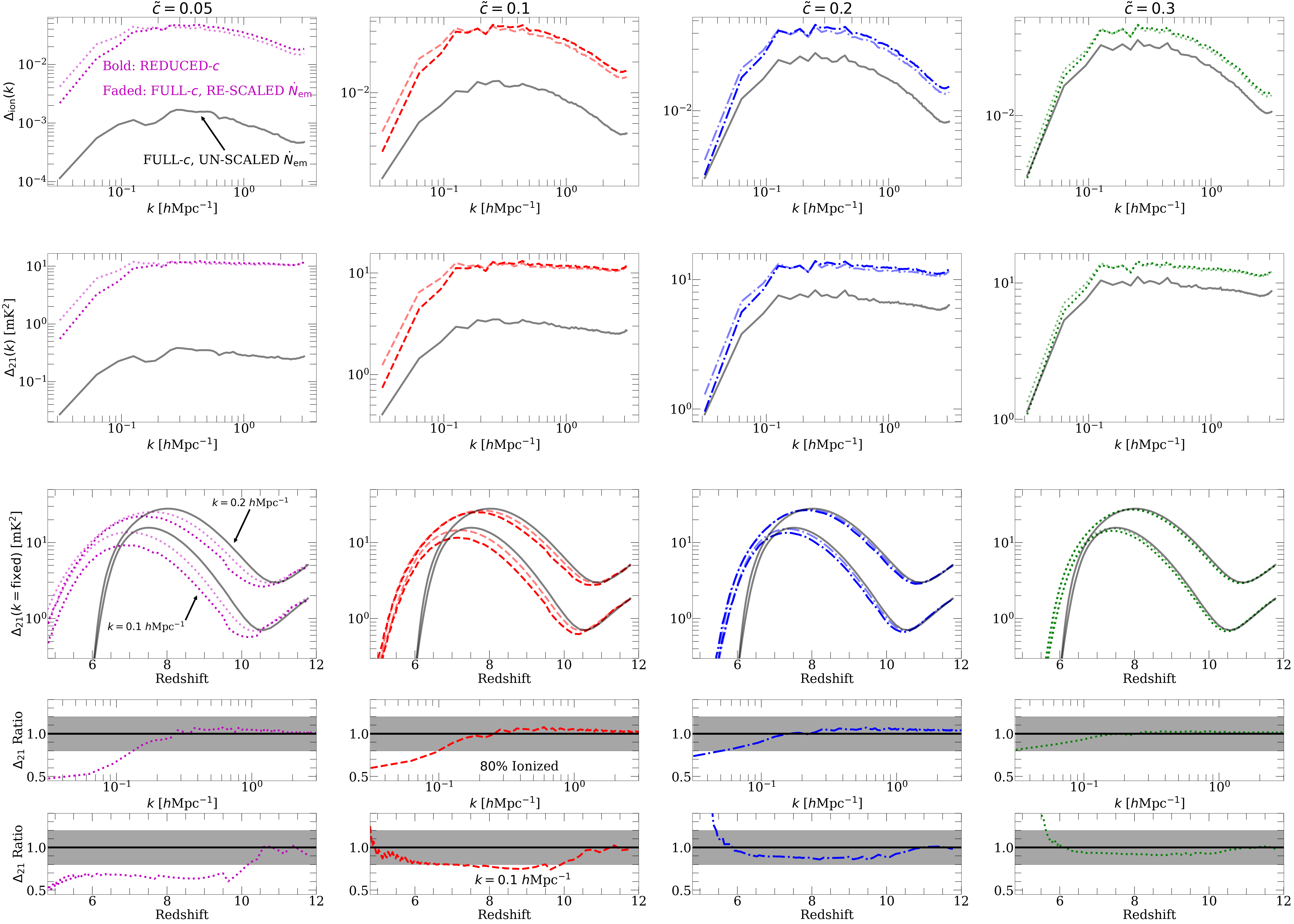}
    \caption{Power spectrum of the ionization field and 21 cm signal for simulations with $\tilde{c} = 0.05$, $0.1$, $0.2$, and $0.3$ (columns from left to right).  {\bf Top Row:} power spectrum of the ionization field, $\Delta_{\rm ion}$, vs. wavenumber $k$ at an ionized fraction of $80\%$.  The bold colored curves show results for the \textsc{reduced-$c$} simulations, and the faded colored curves the \textsc{full-$c$ re-scaled} runs.  The faded gray curves shows (at fixed redshift) the \textsc{full-$c$, un-scaled $\dot{N}_{\rm em}$} runs.  {\bf 2nd Row:} the same, but for the 21 cm power spectrum, $\Delta_{21}$.  The \textsc{full-$c$, un-scaled $\dot{N}_{\rm em}$} results differ dramatically from the other two, reflecting its lower neutral fraction at fixed redshift.  However, the \textsc{reduced-$c$} and \textsc{full-$c$, re-scaled $\dot{N}_{\rm em}$} runs differ in their shape: the former has significantly less power on large scales (small $k$) and slightly more on small-scales (large $k$).  {\bf 3rd Row:} $\Delta_{21}$ vs redshift at $k = 0.2$ and $0.1$ $h$Mpc$^{-1}$ (top and bottom set of curves, respectively).  The bold colored curves are noticeably below the faded ones, reflecting the deficit in large-scale power in the \textsc{reduced-$c$} runs.  {\bf 4th \& 5th Rows:} ratios of the \textsc{reduced-$c$} and \textsc{full-$c$ re-scaled} results in the 2nd and 3rd rows, respectively.  The shaded regions denote $\pm 20\%$ from unity.  In the bottom row, we show the ratio for $k = 0.1$ $h$Mpc$^{-1}$.  Decreasing $\tilde{c}$ increases differences at low $k$, so that for $\tilde{c} = 0.1$ the difference is $\approx 20\%$ at all redshifts at $k = 0.1$ $h$Mpc$^{-1}$.  At $\tilde{c} \geq 0.2$, we find $\lesssim 15\%$ differences at $k = 0.1$ $h$Mpc$^{-1}$, which get smaller at larger $k$.  }
    \label{fig:power_spectrum_ion}
\end{figure*}

Figure~\ref{fig:power_spectrum_ion} quantitatively compares the large-scale fluctuations in the ionization field in all three types of simulations.  In the top row, we show the dimensionless ionization power spectrum $\Delta_{\rm ion}(k)$ at $80\%$ ionized for $\tilde{c} = 0.05$, $0.1$, $0.2$, and $0.3$ (left to right).  The bold colored lines, faded colored lines, and faded gray lines indicate results for \textsc{reduced-$c$}, \textsc{full-$c$, re-scaled $\dot{N}_{\rm em}$}, and \textsc{full-$c$, un-scaled $\dot{N}_{\rm em}$} runs, respectively.  The second row is the same thing, but instead showing the 21 cm power spectrum\footnote{We assume that the spin temperature $T_{\rm S}$ is much greater than the CMB temperature.  }, calculated using the \textsc{tools21cm} package of Ref.~\cite{Giri2018}.  

The power in the \textsc{full-$c$, un-scaled $\dot{N}_{\rm em}$} run is much lower than the others - increasingly so at lower $\tilde{c}$.  This reflects the fact that the \textsc{reduced-$c$} runs are more delayed relative to the \textsc{full-$c$, un-scaled $\dot{N}_{\rm em}$} case the smaller $\tilde{c}$ is.  So, at a fixed neutral fraction in the former, the neutral fraction in the latter decreases for smaller $\tilde{c}$.  The \textsc{full-$c$, re-scaled $\dot{N}_{\rm em}$} sims are much closer to the \textsc{reduced-$c$} results, but differ noticeably in their shape.  The latter have significantly less power at large scales (small $k$), and slightly more power at small scales (large $k$).  This is a direct result of the morphological errors shown in Figure~\ref{fig:ion_field_vis}, with the reduced sizes of the largest bubbles driving the decrease in small-$k$ power.  In the third row, we show $\Delta_{21}$ vs. redshift at fixed $k = 0.2$ and $0.1$ $h$Mpc$^{-1}$ (top and bottom set of curves, respectively), key targets for experiments such as HERA~\citep{HERA2021a,HERA2021b,Berkhout2024}.  The suppression in power is consistent across the entire reionization history, indicating that even early in reionization the RSLA can cause significant errors in ionization morphology.  

The fourth row shows the ratio of the bold and faded colored curves in the second row, with the shaded region denoting $\pm 20\%$.  For $\tilde{c} = 0.3$, the differences between the \textsc{reduced-$c$} and \textsc{full-$c$ re-scaled} runs never exceeds $20\%$ even at the largest scales at $80\%$ ionized.  For lower $\tilde{c}$, the drop in power at small $k$ increases, reaching a factor of $\approx 2$ at the box scale for $\tilde{c} = 0.05$.  For $\tilde{c} = 0.2$ the deviation is still only $\sim 10\%$ at $k = 0.1$ $h$Mpc$^{-1}$, but at $\tilde{c} = 0.1$ this increases to $20\%$, and to $30\%$ for $\tilde{c} = 0.05$.  The bottom row shows the ratio of the $k = 0.1$ $h$Mpc$^{-1}$ curves in the third row.  Here we see that for $\tilde{c} = 0.2$, the differences are never larger than $15\%$ at any redshift.  For $\tilde{c} = 0.1$, the difference exceeds $20\%$ at $z \sim 9$, and for $\tilde{c} = 0.05$, it reaches nearly a factor of $2$.  

We see that the utility of Eq.~\ref{eq:main_result} is limited by the fact that it does not account for the spatial fluctuations in the delay between emission and ionization of ionizing photons in the \textsc{reduced-$c$} simulations.  To summarize, $\Delta_{21}$ at $k \leq 0.1$ $h$Mpc$^{-1}$ can deviate by $20\%$ between the \textsc{reduced-$c$} and \textsc{full-$c$, re-scaled $\dot{N}_{\rm em}$} runs for $\tilde{c} = 0.1$, and by up to $40\%$ for $\tilde{c} = 0.05$.  These differences drop to $\leq 15\%$ for $\tilde{c} = 0.2$ and $\leq 10\%$ for $\tilde{c} = 0.3$.  As such, we caution against applying our method for $\tilde{c} < 0.2$ in contexts where the large-scale reionization morphology is important.  

\subsection{QSO observations at $z \leq 6$}
\label{subsec:qso}

In this section, we apply the same kind of test to observables derived from high-redshift quasar absorption spectra at $z \leq 6$.  These include the Ly$\alpha$ forest, which has recently yielded detailed insights into the tail end of reionization at $5 < z < 6$~\citep{Kulkarni2019,Keating2019,Nasir2020,Qin2021,Bosman2021}, and the ionizing photon mean free path~\citep{Worseck2014,Becker2021,Zhu2023,Gaikwad2023}.  These observables are sensitive not only to ionization morphology, but also to the large-scale spatial fluctuations in the ionizing background $\Gamma_{\rm HI}$ and the IGM temperature, which are affected by the RSLA.    

\subsubsection{Global Observables}
\label{subsubsec:qsoglobal}

The evolution of the mean transmission of the Ly$\alpha$ forest is a tight boundary condition on the end of reionization.  Several studies~\citep{Kulkarni2019,Keating2020,Cain2023,Asthana2024} have ``calibrated'' their reionization simulations to agree by construction with this observable, then asked whether other observables predicted by the simulation agree well with observations.  Here, we will follow this procedure and assess how applicable Eq.~\ref{eq:main_result} is in the context of forest-anchored studies.  We first run a set of \textsc{reduced-$c$} simulations with $\dot{N}_{\rm em}^{\tilde{c}}(z)$ calibrated so that the mean transmission of the Ly$\alpha$ forest matches the recent observations of Ref.~\cite{Bosman2021}.  Note that unlike in the previous section, this procedure yields a different $\dot{N}_{\rm em}^{\tilde{c}}$ for each value of $\tilde{c}$, since we are requiring all \textsc{reduced-$c$} runs to match the same observable.  For each $\tilde{c}$, we apply Eq.~\ref{eq:main_result} to $\dot{N}_{\rm em}^{\tilde{c}}$ and run \textsc{full-$c$, re-scaled $\dot{N}_{\rm em}$} simulations using $c = 1$.  In some plots, we will also include results from \textsc{full-$c$, un-scaled $\dot{N}_{\rm em}$} runs.  We will judge the accuracy of Eq.~\ref{eq:main_result} in the same way as in the previous section - by how well the \textsc{reduced-$c$} and \textsc{full-$c$, re-scaled $\dot{N}_{\rm em}$} runs agree with each other.  

\begin{figure}
    \centering
    \includegraphics[scale=0.18]{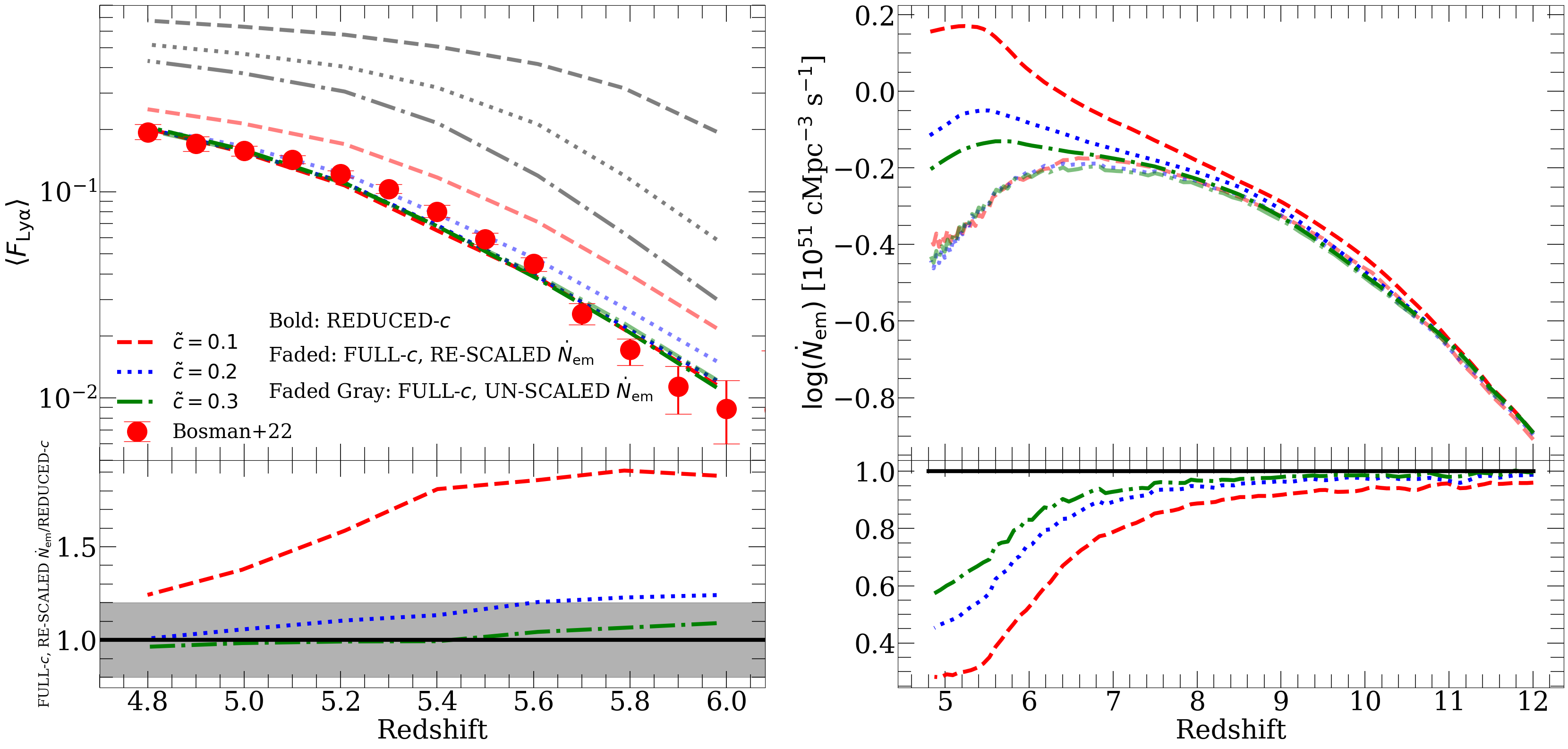}
    \caption{Tests with simulations calibrated to match the Ly$\alpha$ forest at $z = 6$.  {\bf Upper Left:} $\langle
    F_{\rm Ly\alpha} \rangle$ vs. redshift for all three types of simulations described in Table~\ref{tab:terminology} for $\tilde{c} = 0.1$, $0.2$, and $0.3$.  The formatting of the curves is the same used in Figure~\ref{fig:power_spectrum_ion} (see text and legend).  By construction, all the \textsc{reduced-$c$} results agree with each other and the measurements from Ref.~\cite{Bosman2021}.  The \textsc{full-$c$, re-scaled $\dot{N}_{\rm em}$}, $\tilde{c} = 0.3$ run is indistinguishable from these.  The $\tilde{c} = 0.2$ run is slightly above the others, but still in reasonable agreement with the measurements.  However, the $\tilde{c} = 0.1$ run diverges significantly from the measurements.  All the \textsc{full-$c$, un-scaled $\dot{N}_{\rm em}$} have too much transmission, since they end reionization earlier than the \textsc{reduced-$c$} runs.  {\bf Bottom Left:} ratio of $\langle F_{\rm Ly\alpha} \rangle$ in the \textsc{full-$c$, re-scaled $\dot{N}_{\rm em}$} and \textsc{reduced-$c$} runs, with the shaded region denoting $\pm 20\%$.  We find $\lesssim 5\%$ agreement for $\tilde{c} = 0.3$ and $\lesssim 20\%$ agreement for $\tilde{c} = 0.2$, but for $\tilde{c} = 0.1$, the simulations disagree by a factor of $1.3-2$.  {\bf Upper Right:}  $\dot{N}_{\rm em}$ for the \textsc{reduced-$c$} (bold curves) and \textsc{full-$c$, re-scaled $\dot{N}_{\rm em}$} (faded curves) simulations.  As previously mentioned, the \textsc{reduced-$c$} runs have different $\dot{N}_{\rm em}$ since they are all separately calibrated to match the forest.  The \textsc{full-$c$, re-scaled $\dot{N}_{\rm em}$} runs have similar $\dot{N}_{\rm em}$, which is expected since they use the full speed of light and are supposed to match the same observable, $\langle F_{\rm Ly\alpha}\rangle$.  {\bf Bottom Right:} ratio of $\dot{N}_{\rm em}$ in the \textsc{full-$c$, re-scaled $\dot{N}_{\rm em}$} and \textsc{reduced-$c$} runs.  They are similar early in reionization but diverge by up to a factor of $3$ by its end.  }
    \label{fig:forest_flux_plot} 
\end{figure}

We first compare the mean transmission of the Ly$\alpha$ forest, $\langle F_{\rm Ly\alpha} \rangle$, with measurements from Ref.~\cite{Bosman2021} in the top left panel of Figure~\ref{fig:forest_flux_plot}.  The bold colored lines show \textsc{reduced-$c$} runs for $\tilde{c} = 0.1$, $0.2$, and $0.3$.  Faded colored lines (with matching line styles) show the \textsc{full-$c$, re-scaled $\dot{N}_{\rm em}$} results, and faded gray lines (again, with matching line styles) show the \textsc{full-$c$, un-scaled $\dot{N}_{\rm em}$} results.  The bottom panel shows the ratios of the  \textsc{full-$c$, re-scaled $\dot{N}_{\rm em}$} and \textsc{reduced-$c$} results (with the shaded line again denoting $\pm 20\%$ from unity). 
 All the \textsc{reduced-$c$} results agree well with the measurements and each other by construction.  The \textsc{full-$c$, re-scaled $\dot{N}_{\rm em}$} run with $\tilde{c} = 0.3$ is indistinguishable from its \textsc{reduced-$c$} counterpart.  For $\tilde{c} = 0.2$, $\langle F_{\rm Ly\alpha}\rangle$ for the \textsc{full-$c$, re-scaled $\dot{N}_{\rm em}$} run slightly above the measurements, but is still within at most $25\%$ of its \textsc{reduced-$c$} counterpart.  However, for $\tilde{c} = 0.1$, the \textsc{full-$c$, re-scaled $\dot{N}_{\rm em}$} run diverges by as much as a factor of $2$.  All three of the \textsc{full-$c$, un-scaled $\dot{N}_{\rm em}$} lie well above the measurements, since their reionization histories end too early.  

In the upper right panel, we show the emissivities for the \textsc{reduced-$c$} and \textsc{full-$c$, re-scaled $\dot{N}_{\rm em}$} runs ($\dot{N}_{\rm em}^{\tilde{c}}$ and $\dot{N}_{\rm em}^{c}$, respectively).  The bottom panel shows their ratio.  We see that $\dot{N}_{\rm em}^{\tilde{c}}$ is quite different for different values of $\tilde{c}$, since all runs were calibrated to agree with the same observable.  On the other hand, $\dot{N}_{\rm em}^{c}$ is quite similar between the different values of $\tilde{c}$.  This is indeed to be expected if Eq.~\ref{eq:main_result} is accurate, since each $\dot{N}_{\rm em}^{c}$ is supposed to match the same observable, $\langle F_{\rm Ly\alpha} \rangle$, with the full speed of light.  In the lower right panel, we see that the $\dot{N}_{\rm em}^{c}$ and $\dot{N}_{\rm em}^{\tilde{c}}$ are similar early in reionization, but differ by a factor of $2-3$ by its end.  

\begin{figure}
    \centering
    \includegraphics[scale=0.215]{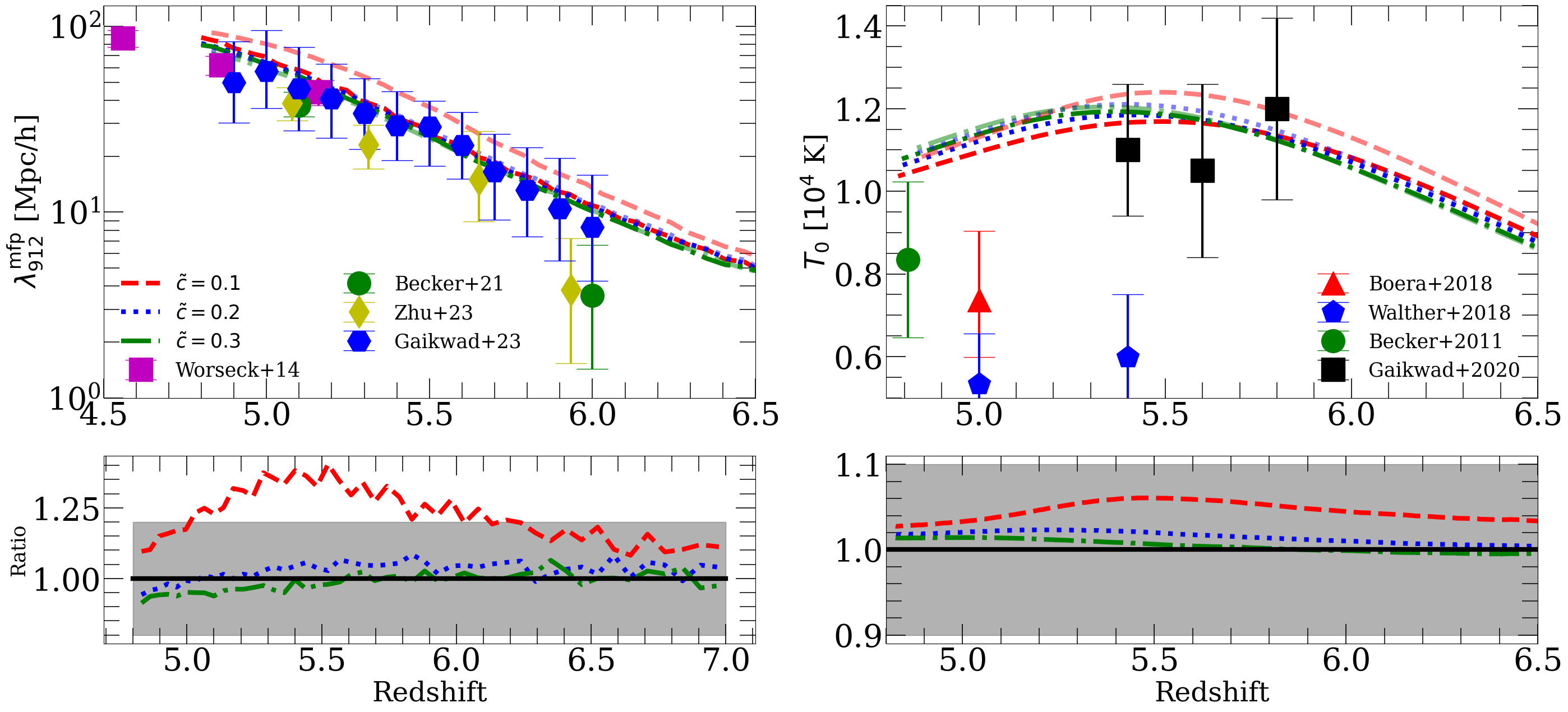}
    \caption{The same as Figure~\ref{fig:forest_flux_plot}, but showing $\lambda_{\rm mfp}^{912}$ (left) and $T_0$ (right).  To avoid cluttering the plot, we omit the \textsc{full-$c$, un-scaled $\dot{N}_{\rm em}$} results.  For both $\tilde{c} = 0.3$ and $0.2$, we see agreement to within a few percent in both quantities between the \textsc{reduced-$c$} and \textsc{full-$c$, re-scaled $\dot{N}_{\rm em}$} runs.  However, for $\tilde{c} = 0.1$, two diverge by up to $\approx 35\%$ for $\lambda_{\rm mfp}^{912}$ and $\approx 6\%$ for $T_0$.  These results further confirm our earlier observations that our method should only be applied for $\tilde{c} \geq 0.2$.  }
    \label{fig:mfp_temp_global}
\end{figure}

In Figure~\ref{fig:mfp_temp_global}, we make the same comparison for the lyman limit mean free path\footnote{We calculate $\lambda_{\rm mfp}$ using the method described in appendix C of Ref.~\cite{Chardin2015}.  } ($\lambda_{912}^{\rm mfp}$, MFP, left) and the IGM temperature at mean density $T_0$ (right).  To avoid cluttering the plot, we omit results from the \textsc{full-$c$, un-scaled $\dot{N}_{\rm em}$} runs.  We also include several sets of observations for comparison (see references in caption).  For $\tilde{c} = 0.2$ and $0.3$, the \textsc{reduced-$c$} and \textsc{full-$c$, re-scaled $\dot{N}_{\rm em}$} agree to within a few percent in both $\lambda_{\rm mfp}^{912}$ and $T_0$, as the bottom panels show.  For $\tilde{c} = 0.1$, the disagreement is much larger - up to $35\%$ for $\lambda_{\rm mfp}^{912}$ and $\approx 6\%$ for $T_0$.  Thus, the trend here is qualitatively similar to that observed in Figures~\ref{fig:power_spectrum_ion} and~\ref{fig:forest_flux_plot} - namely, that Eq.~\ref{eq:main_result} gives a reasonable match between \textsc{reduced-$c$} and \textsc{full-$c$, re-scaled $\dot{N}_{\rm em}$} simulations for $\tilde{c} = 0.2$, but not for $\tilde{c} = 0.1$.  This further confirms that our approach can only be reliably used down to $\tilde{c} = 0.2$.  

\subsubsection{Large-scale fluctuations}
\label{subsubsec:large_scale_flucs}

We saw in \S\ref{subsubsec:ion_morphology} and \S\ref{subsubsec:qsoglobal} that the applicability of Eq.~\ref{eq:main_result} is likely limited to $\tilde{c} \gtrsim 0.2$ by spatially-varying time-delay effects caused by the RSLA on large scales.  In this section, we study this effect in more detail in the context of the Ly$\alpha$ forest.  At the end of reionization, large-scale fluctuations in forest properties are set by three quantities that the RSLA affects: the ionization morphology~\cite{Kulkarni2019}, large-scale fluctuations in $\Gamma_{\rm HI}$~\cite{Davies2016}, and fluctuations in temperature~\cite{DAloisio2015}.  We will focus on the latter two in this section.   

\begin{figure}
    \centering
    \includegraphics[scale=0.16]{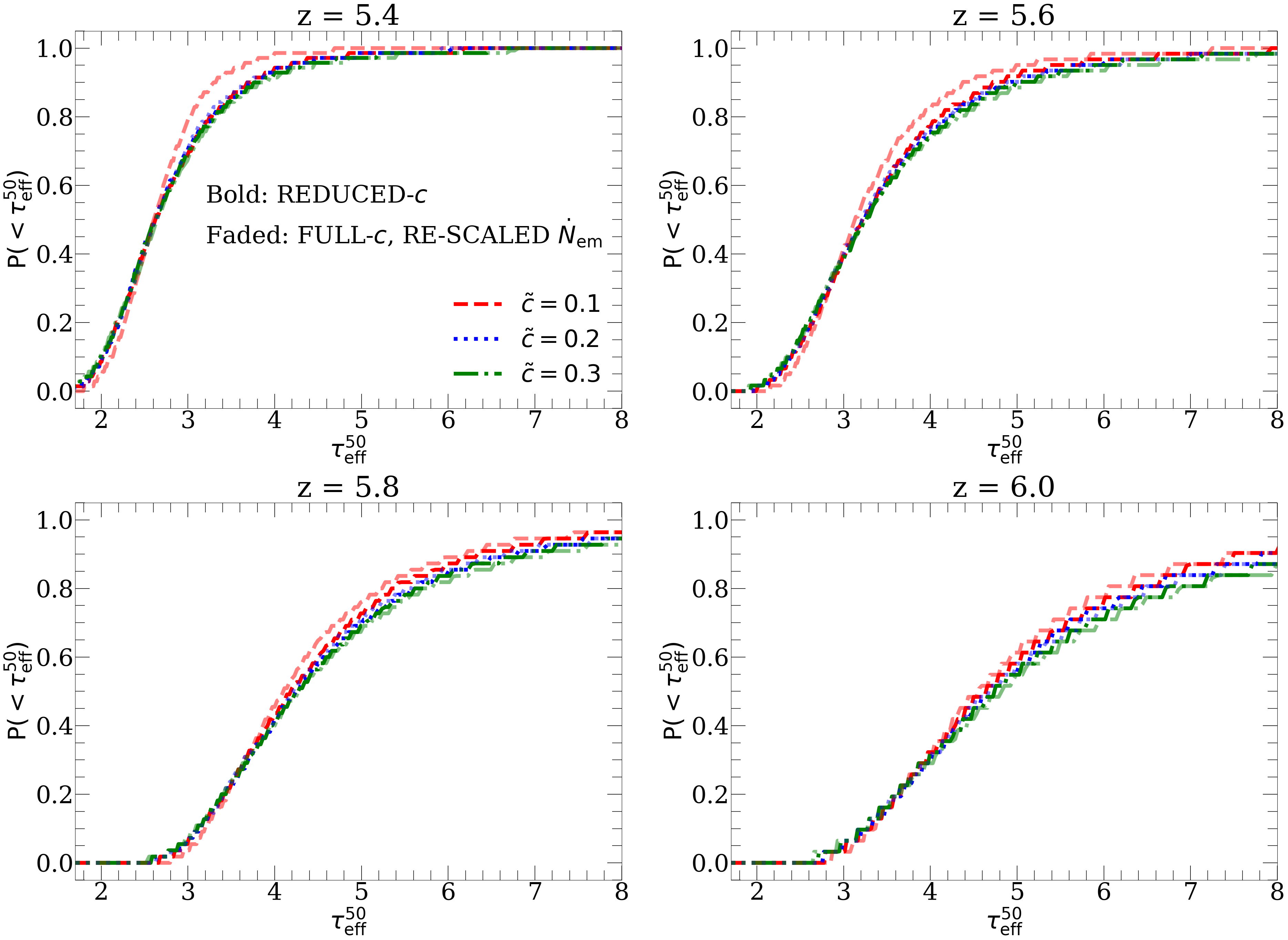}
    \caption{Distribution of Ly$\alpha$ forest effective optical depths over $50$ $h^{-1}$Mpc segments of the forest, $P(<\tau_{\rm eff}^{50})$, at $z = 5.4$, $5.6$, $5.8$, and $6.0$.  The line styles and colors have the same meanings as in Figure~\ref{fig:mfp_temp_global}.  At $z = 5.8$ and $6$, when the neutral fraction is $> 10\%$, the \textsc{reduced-$c$} results differ from each other at the few percent level.  This is unsurprising since, as we showed earlier, they differ in large-scale ionization morphology.  The $\tilde{c} = 0.2$ and $0.3$ \textsc{full-$c$, re-scaled $\dot{N}_{\rm em}$} agree with their \textsc{reduced-$c$} counterparts to within a few percent, but this deviation grows noticeably larger for $\tilde{c} = 0.1$, especially at $z = 5.4$ and $5.6$.  This suggests that the effect of the RSLA on fluctuations in $\Gamma_{\rm HI}$ and $T$ may be affecting these differences when the neutral fraction is $< 10\%$.  }
    \label{fig:taueff_reduced_c}
\end{figure}

In Figure~\ref{fig:taueff_reduced_c}, we show the cumulative distribution function (CDF) of $\tau_{\rm eff}^{50}$, $P(<\tau_{\rm eff}^{50})$, the effective optical depth over $50$ $h^{-1}$Mpc segments of the forest, at $z = 5.4$, $5.6$, and $5.8$, and $6.0$.  This statistic has been used in a number of works to argue that reionization must have ended later than $z = 6$~\cite{Kulkarni2019,Keating2019,Keating2020,Nasir2020,Choudhury2021,Qin2021}.  The format of the curves is the same as that in Figure~\ref{fig:mfp_temp_global}.  All the \textsc{reduced-$c$} runs, and the \textsc{full-$c$, re-scaled $\dot{N}_{\rm em}$} runs with $\tilde{c} = 0.2$ and $0.3$ (faded blue-dotted and green dot-dashed curves) mutually agree to within a few percent at all redshifts.  We do see few-percent level differences between different \textsc{reduced-$c$} runs at $z = 6$ and $5.8$, which get smaller at $z = 5.6$ and $5.4$.  This is unsurprising, since we have already shown that the \textsc{reduced-$c$} runs differ from each other in large-scale ionization morphology (Figures~\ref{fig:ion_field_vis}-\ref{fig:power_spectrum_ion}), which also affects the $\tau_{\rm eff}^{50}$ CDF when the IGM is still partially neutral.  We also see that the \textsc{full-$c$, re-scaled $\dot{N}_{\rm em}$} $\tilde{c} = 0.1$ run differs significantly from the others, especially at $z = 5.6$ and $5.4$, when the neutral fraction is $< 10\%$.  This suggests that the effect of the RSLA on large-scale fluctuations in $\Gamma_{\rm HI}$ and/or $T$ may also play a significant role in driving these differences.  

\begin{figure}
    \centering
    \includegraphics[scale=0.285]{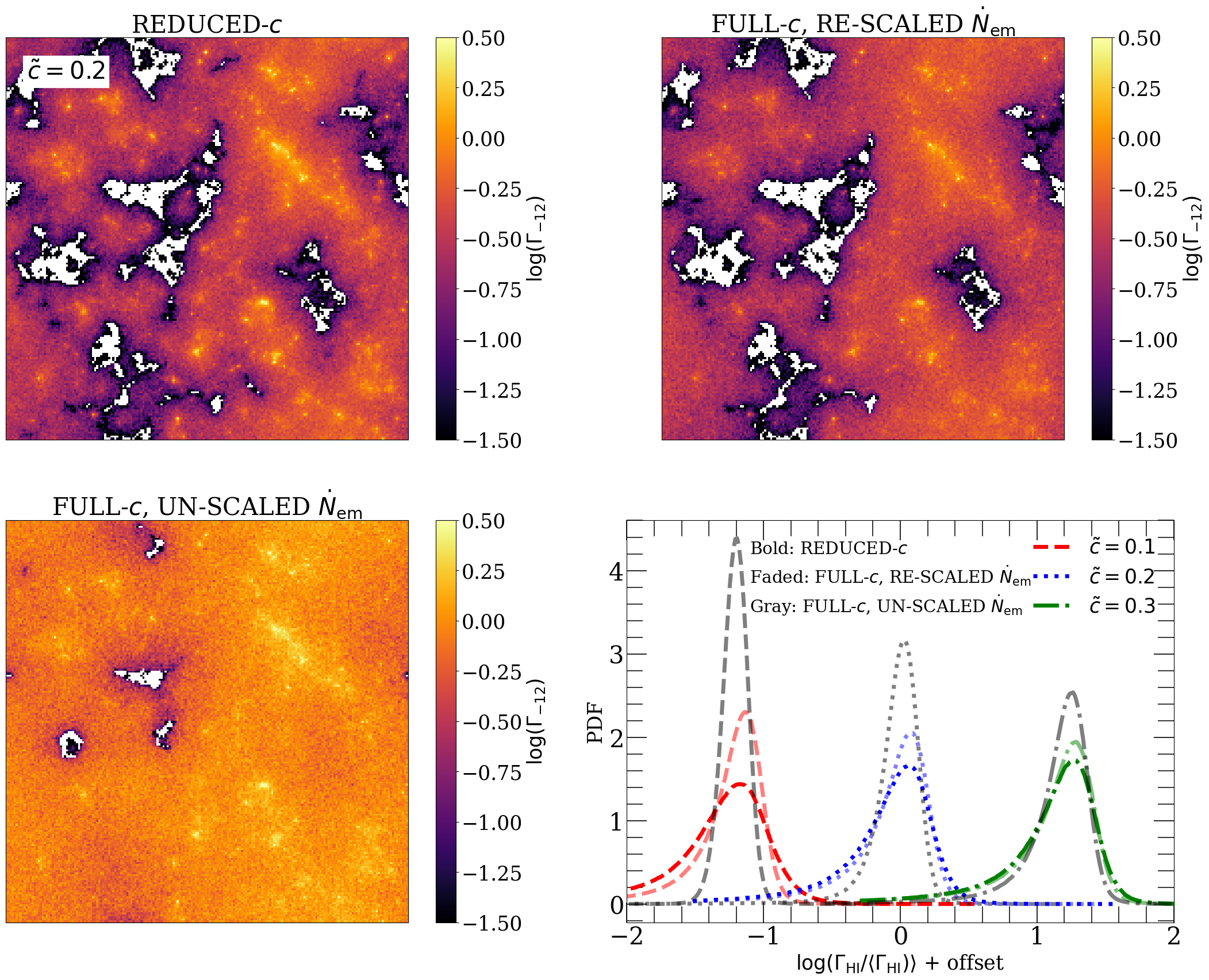}
    \caption{Visualization of the effect of the RSLA on large-scale fluctuations in $\Gamma_{\rm HI}$ (in units of $10^{-12}$ s$^{-1}$).  We show maps of the \textsc{reduced-$c$}, \textsc{full-$c$, re-scaled $\dot{N}_{\rm em}$}, and \textsc{full-$c$, un-scaled $\dot{N}_{\rm em}$} runs in the upper left, upper right, and lower left panels, respectively, for $\tilde{c} = 0.2$.  The \textsc{full-$c$, un-scaled $\dot{N}_{\rm em}$} run has a higher mean $\Gamma_{\rm HI}$, fewer neutral islands (white regions), and weaker spatial fluctuations in $\Gamma_{\rm HI}$, since reionization ends earlier in that run than in the others.  The \textsc{reduced-$c$} and \textsc{full-$c$, re-scaled $\dot{N}_{\rm em}$} runs are in reasonably good visual agreement.  However, the former has noticeably larger $\Gamma_{\rm HI}$ fluctuations, with the bright (faint) regions being brighter (fainter) than in the \textsc{full-$c$, re-scaled $\dot{N}_{\rm em}$} run.  In the lower right, we plot log-space PDFs of $\Gamma_{\rm HI}$ normalized by its mean.  We show results for all three values of $\tilde{c}$, with the same line styles and colors used in Figure~\ref{fig:forest_flux_plot}.  We have offset the $\tilde{c} = 0.1$ ($0.3$) PDFs by $1.2$ dex to the left (right) for clarity.  In all three cases, the PDFs are much narrower in the \textsc{full-$c$, un-scaled $\dot{N}_{\rm em}$} runs than in the others.  As $\tilde{c}$ decreases, the PDFs in the \textsc{reduced-$c$} become grow wider than their \textsc{full-$c$, re-scaled $\dot{N}_{\rm em}$} counterparts, owing to time-delay effects (see text). 
 }
    \label{fig:gamma_vis}
\end{figure}

We visualize the effect of the RSLA on large-scale fluctuations in $\Gamma_{\rm HI}$ in Figure~\ref{fig:gamma_vis}.  We show maps of $\Gamma_{\rm HI}$ (in units of $10^{-12}$ s$^{-1}$) at $z = 5.65$, when the universe is $\approx 10\%$ neutral in the \textsc{reduced-$c$} and \textsc{full-$c$, re-scaled $\dot{N}_{\rm em}$} runs.  The top left, top right, and lower left panels show maps for the \textsc{reduced-$c$}, \textsc{full-$c$, re-scaled $\dot{N}_{\rm em}$}, and \textsc{full-$c$, un-scaled $\dot{N}_{\rm em}$} runs, respectively, for $\tilde{c} = 0.2$.  We see that the \textsc{full-$c$, un-scaled $\dot{N}_{\rm em}$} run has a much higher mean $\Gamma_{\rm HI}$ and weaker $\Gamma_{\rm HI}$ fluctuations than the other two, owing to its earlier end to reionization.  Compared to this run, the \textsc{reduced-$c$} and \textsc{full-$c$, re-scaled $\dot{N}_{\rm em}$} runs are in relatively good agreement.  However, the spatial fluctuations in $\Gamma_{\rm HI}$ are noticeably stronger in the \textsc{reduced-$c$} run.  Once again, this is due to time-delay effects caused by the RSLA.  Because photons are delayed longest in reaching the edges of the largest ionized regions, the differences in $N_{\gamma}$ between them and the smaller bubbles are amplified.  In addition, photons far from sources were emitted at a retarded time $t_{\rm ret} = R/\tilde{c}$, where $R$ is the ionized bubble size.  Again, these effects are accounted for only on average by the re-scaling of Eq.~\ref{eq:main_result}, which results in larger spatial $\Gamma_{\rm HI}$ fluctuations in the \textsc{reduced-$c$} runs than in their \textsc{full-$c$, re-scaled $\dot{N}_{\rm em}$} counterparts.  

In the lower right panel, we show the PDFs of $\Gamma_{\rm HI}$, normalized by its mean (in log space).  Here, we show results for all three values of $\tilde{c}$, using the same color and line style convention adopted in Figure~\ref{fig:forest_flux_plot}.  To make the plot readable, we have offset the $\tilde{c} = 0.1$ ($0.3$) PDFs to the left (right) by $1.2$ dex.  In all cases, the gray curves are much narrower than the others, reflecting the reduced $\Gamma_{\rm HI}$ fluctuations in the \textsc{full-$c$, un-scaled $\dot{N}_{\rm em}$} runs.  As $\tilde{c}$ decreases, the PDFs of the \textsc{reduced-$c$} runs grow noticeably wider than those of their \textsc{full-$c$, re-scaled $\dot{N}_{\rm em}$} counterparts, with the difference becoming very significant for $\tilde{c} = 0.1$.  These differences contribute to those in $P(< \tau_{\rm eff}^{50})$ seen in Figure~\ref{fig:taueff_reduced_c}.

\begin{figure}
    \centering
    \includegraphics[scale=0.285]{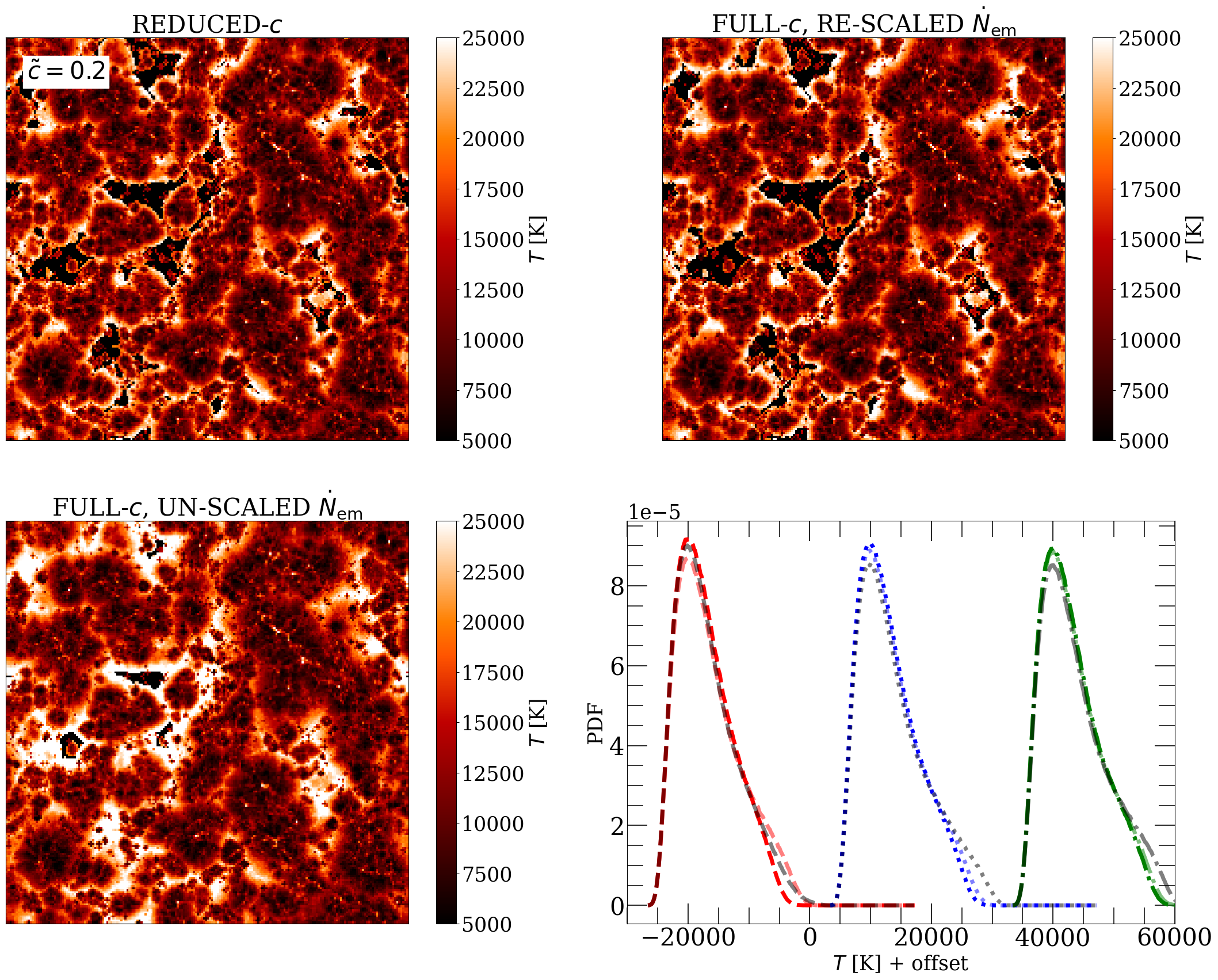}
    \caption{The same as Figure~\ref{fig:gamma_vis}, but for the IGM gas temperature.  The main difference between the three maps is the strength of the temperature enhancements near neutral islands, where gas has been recently reionized and is hottest.  The \textsc{full-$c$, un-scaled $\dot{N}_{\rm em}$} run has the hottest temperatures in the voids, since reionization ends earliest and most rapidly.  The \textsc{reduced-$c$} run, by contrast, has the weakest $T$ fluctuations.  This is because the slowed growth of the largest ionized regions (see \S\ref{subsubsec:ion_morphology}) reduces the speed of the fastest I-fronts, which lowers $T_{\rm reion}$ around the largest bubbles.  In the lower right panel, we see a significant difference between \textsc{reduced-$c$} and \textsc{full-$c$, re-scaled $\dot{N}_{\rm em}$} runs for $\tilde{c} = 0.1$ at the high-$T$ end of the PDF, which explains the differences in $T_0$ seen in Figure~\ref{fig:mfp_temp_global}.  }
    \label{fig:temp_vis}
\end{figure}

Figure~\ref{fig:temp_vis} shows the same thing as Figure~\ref{fig:gamma_vis}, but for IGM temperature.  In the lower right panel, we offset the $\tilde{c} = 0.1$ and $0.3$ results by $\pm 30000$ K for clarity.  The voids, which have ionized most recently, are hottest in the \textsc{full-$c$, un-scaled $\dot{N}_{\rm em}$} run, which ends reionization the earliest and most rapidly.  The \textsc{reduced-$c$} run displays the smallest such $T$ enhancements.  The fastest I-fronts around the largest ionized regions, which produce the largest $T_{\rm reion}$~\citep{DAloisio2019,Zeng2021}, grow move more slowly than they should in the \textsc{reduced-$c$} runs for aforementioned reasons.  We see in the lower right panel that the difference between the \textsc{reduced-$c$} and \textsc{full-$c$, re-scaled $\dot{N}_{\rm em}$} in the high-$T$ tail of the PDF becomes significant for $\tilde{c} = 0.1$, explaining the corresponding differences in $T_0$ in Figure~\ref{fig:mfp_temp_global}.  

We have repeated the tests in this section, in whole or in part, for several variations of the properties of the ionized sources and the IGM assumed in this work.  We have first varied the clustering of ionizing sources assumed in our fiducial scenario, which assumes that the ionizing emissivity of halos scales with their UV luminosity.  We have tested models wherein the bulk of the ionizing photons are emitted by the only the least massive (least clustered) and most massive (most clustered) halos.  We have also applied our method to multi-frequency RT simulations, for which we show results in the next section.  Lastly, we tested models that assume more sub-grid recombinations than the Reference model of Ref.~\cite{Cain2023}, and we have varied the reionization history.  We find that our results remain essentially the same for all of these variations of our fiducial scenario - namely, that Eq.~\ref{eq:main_result} is accurate (to $20\%$ or better) for $\tilde{c} = 0.2$, but that using $\tilde{c} = 0.1$ leads to much larger errors.  

These results further corroborate our conclusions from \S\ref{subsec:ion_hist_morph}.  We find that, for $\tilde{c} \geq 0.2$, \textsc{reduced-$c$} simulations calibrated to match the Ly$\alpha$ forest match well the properties of their \textsc{full-$c$, re-scaled $\dot{N}_{\rm em}$} counterparts.  This confirms that Eq.~\ref{eq:main_result} can be applied to simulations using the RSLA with $\tilde{c} \geq 0.2$.  However, for $\tilde{c} = 0.1$, we see significant differences between the \textsc{reduced-$c$} and \textsc{full-$c$, re-scaled $\dot{N}_{\rm em}$} runs.  These can largely be attributed to position-dependent time-delay effects that arise from using a reduced speed of light, for which Eq.~\ref{eq:main_result} does not account.  As such, for applications in which the morphology of reionization and/or quasar-based observables at $z \leq 6$ are important, we caution against applying our re-scaling method for $\tilde{c}$ smaller than $0.2$.   

\section{Generalization to multi-frequency simulations}
\label{sec:multi-freq}

In the previous section, we tested Eq.~\ref{eq:main_result} in mono-chromatic simulations.  Here, we will generalize Eq.~\ref{eq:main_result} to apply to multi-frequency simulations and demonstrate that the method works just as well in this case, but with one key additional caveat.  For our purposes, the main effect of including multi-frequency RT is to change the temperature and ionizing background in the ionized IGM~\citep{Cain2023}, but not the morphology of ionized regions.  As such, we will only show our Ly$\alpha$ forest-focused tests (\S\ref{subsec:qso}) in this section.  

A straightforward multi-frequency generalization of Eq.~\ref{eq:main_result} is to simply apply the arguments of \S\ref{subsec:fix} in each frequency bin separately.  Then we have
\begin{equation}
    \label{eq:main_result_nu}
    \dot{N}_{\rm em}^{c}(t,\nu) = \dot{N}_{\rm em}^{\tilde{c}}(t,\nu) - \dot{N}_{\gamma}^{\tilde{c}}(t,\nu) \left(1 - \frac{\tilde{c}}{c}\right)
\end{equation}
A subtlety of Eq.~\ref{eq:main_result_nu} is that in general, $\dot{N}_{\rm em}^{\tilde{c}}(t,\nu)$ and $\dot{N}_{\gamma}^{\tilde{c}}(t,\nu)$ do not share the same shape in frequency space.  Indeed, this is the case during reionization because the absorption cross-section of HI is frequency-dependent, which results in hardening of the radiation field by the IGM.  As such $\dot{N}_{\rm em}^{c}(t,\nu)$ will have a softer ionizing spectrum than $\dot{N}_{\rm em}^{\tilde{c}}(t,\nu)$, since $\dot{N}_{\gamma}^{\tilde{c}}(t,\nu)$ has a harder spectrum.  This complicates somewhat the physical interpretation of the re-scaling.  

\begin{figure}
    \centering
    \includegraphics[scale=0.18]{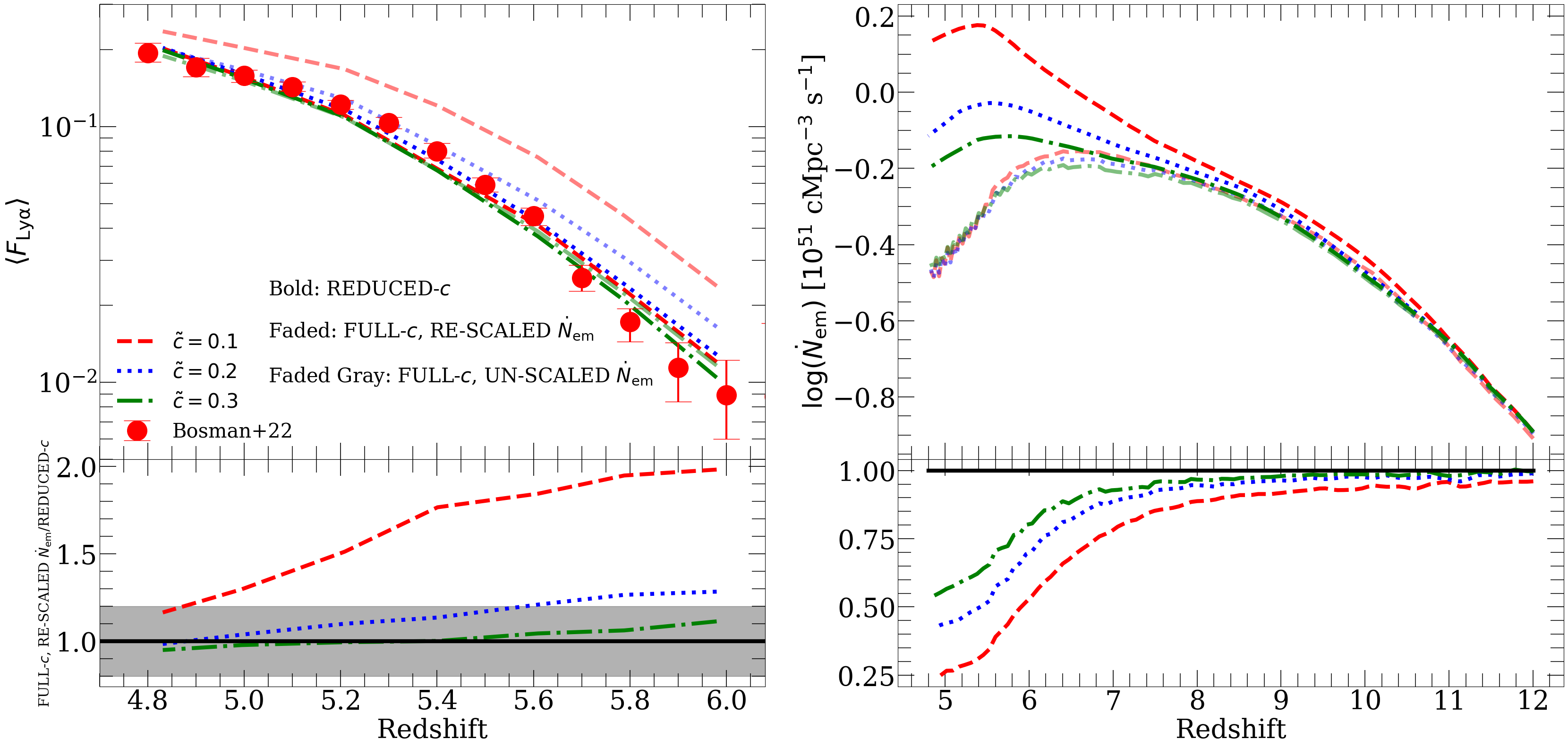}
    \caption{Same as Figure~\ref{fig:forest_flux_plot}, but showing results for our tests with multi-frequency RT, using Eq.~\ref{eq:main_result_nu} to re-scale the emissivity.  We find very similar results in all panels to those in Figure~\ref{fig:forest_flux_plot}, showing that our method works nearly as well for multi-frequency RT simulations as for single-frequency ones.  See text for details.  }
    \label{fig:forest_flux_plot_mfreq}
\end{figure}

In this section, we use the same setup used for multi-frequency RT simulations described in Ref.~\cite{Cain2023}, assuming that the intrinsic spectrum of ionizing sources is a power law of the form $J_{\nu} \propto \nu^{-\alpha}$ with $\alpha = 1.5$.  We use $5$ frequency bins, with central frequencies chosen such that (initially) all bins contain an equal fraction of the emitted ionizing photons.  The condition for choosing the frequency bins is that the average HI-ionizing cross-section, $\overline{\sigma_{\rm HI}}$, should be the same as that of an $\alpha = 1.5$ power law.  As in the previous section, we ran \textsc{reduced-$c$} simulations with $\tilde{c} = 0.1$, $0.2$, and $0.3$, and we apply Eq.~\ref{eq:main_result_nu} to run their \textsc{full-$c$, re-scaled $\dot{N}_{\rm em}$} counterparts.  In Figure~\ref{fig:forest_flux_plot_mfreq}, we show $\langle F_{\rm Ly\alpha}\rangle$ in the same format as Figure~\ref{fig:forest_flux_plot} for our  multi-frequency tests, except that we omit the \textsc{full-$c$, un-scaled $\dot{N}_{
\rm em}$} results.  We find results very similar to those in Figure~\ref{fig:forest_flux_plot}.  The mean flux in the \textsc{reduced-$c$} and \textsc{full-$c$, re-scaled $\dot{N}_{\rm em}$} runs remains within $10\%$ for $\tilde{c} = 0.3$, and within $30\%$ or better for $\tilde{c} = 0.2$.  For $\tilde{c} = 0.1$, we again find differences as large as a factor of $2$ in $\langle F_{\rm Ly\alpha}\rangle$.  In the right column, we see that $\dot{N}_{\rm em}^{\tilde{c}}$ and $\dot{N}_{\rm em}^{c}$, and the relationship between them, are very similar in the multi-frequency runs as in the mono-chromatic case.  This shows that our method works just as well when applied to multi-frequency simulations with respect to the mean Ly$\alpha$ forest transmission.  

\begin{figure}
    \centering
    \includegraphics[scale=0.16]{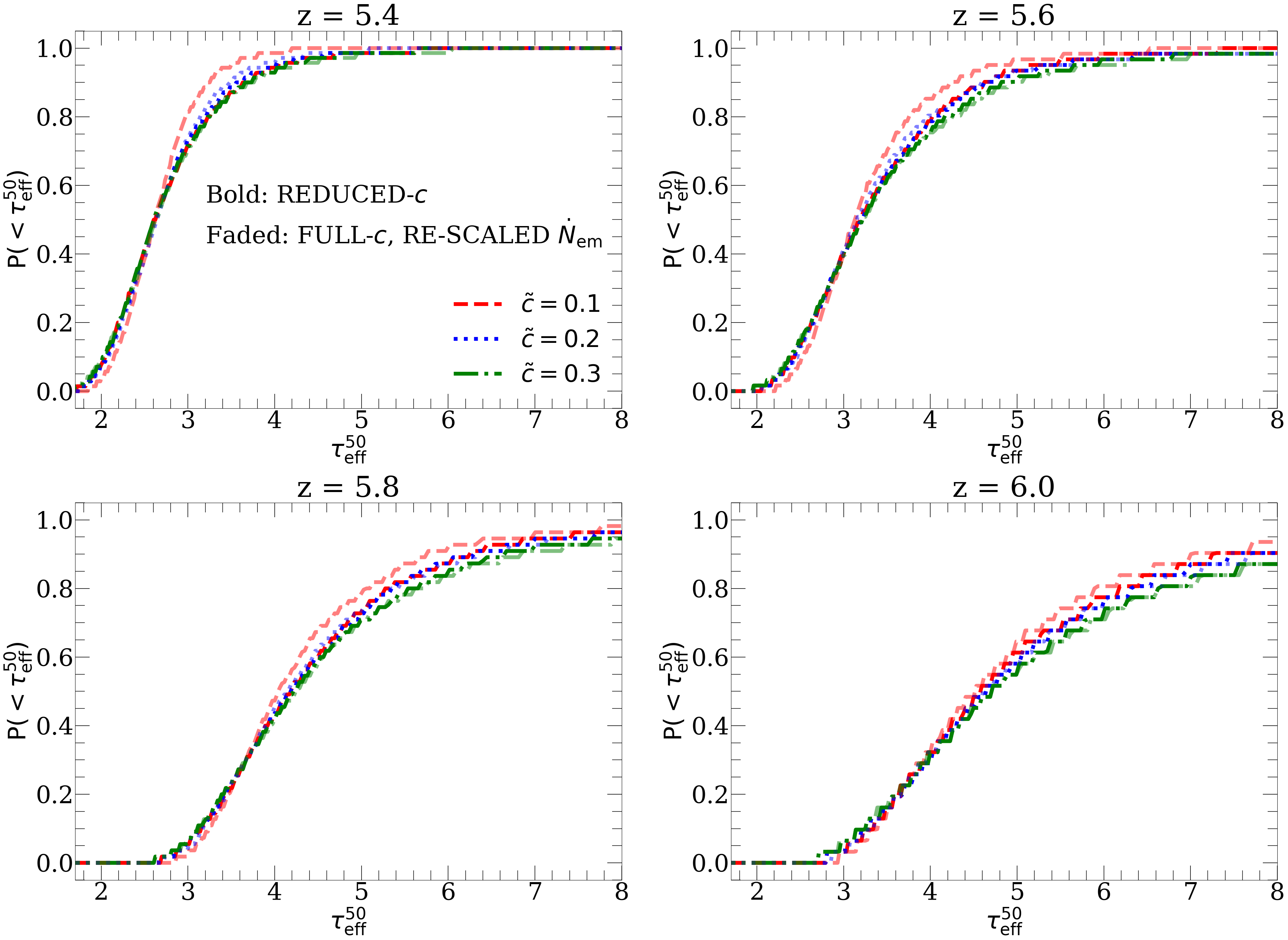}
    \caption{Same as Figure~\ref{fig:taueff_reduced_c}, but showing $P(< \tau_{\rm eff}^{50})$ for our multi-frequency RT tests.  We find results very similar to those of the mono-chromatic runs, showing that our re-scaling procedure works well in the multi-frequency case.  See text for details.  }
    \label{fig:taueff_reduced_c_mfreq}
\end{figure}

In Figure~\ref{fig:taueff_reduced_c_mfreq}, we show $P(<\tau_{\rm eff}^{50})$, in the same format as Figure~\ref{fig:taueff_reduced_c}, for our multi-frequency tests.  Again, we see results very similar to those in Figure~\ref{fig:taueff_reduced_c}.  The differences between the bold and faded red curves are slightly larger than they are in Figure~\ref{fig:taueff_reduced_c}, suggesting that spatial fluctuations in the spectrum of the ionizing background may be slightly worsening the time-delay effects discussed in \S\ref{subsubsec:large_scale_flucs}.  However, this effect is not significant enough to change the level of accuracy obtained with $\tilde{c} = 0.2$ and $0.3$, further validating that for these values of $\tilde{c}$, our method can safely be applied to simulations with multi-frequency RT.  

\begin{figure}
    \centering
    \includegraphics[scale=0.145]{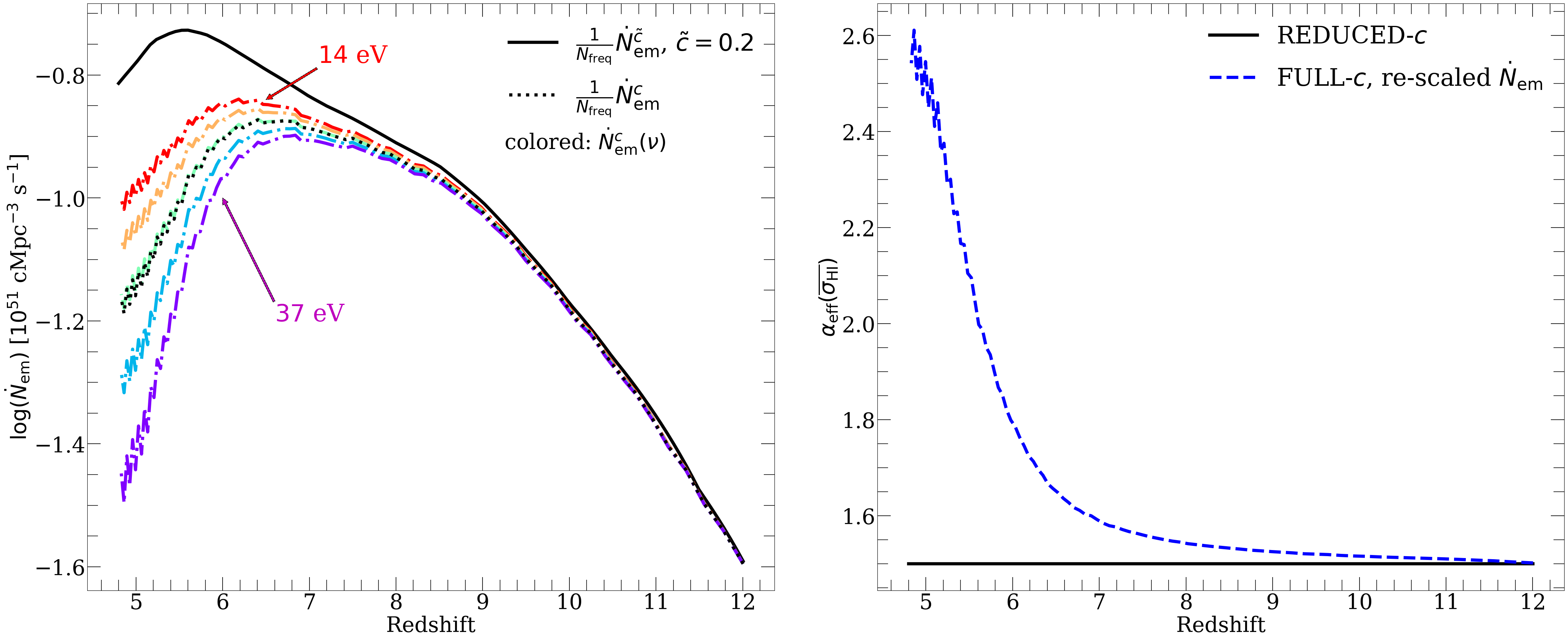}
    \caption{Illustration of how our re-scaling procedure changes the intrinsic spectrum of the sources when applied to multi-frequency RT simulations.  {\bf Left}: Ionizing emissivity in each bin in the \textsc{reduced-$c$} run (black solid) compared to the spectrum in the \textsc{full-$c$, re-scaled $\dot{N}_{\rm em}$} run.  The black dotted line shows the average fraction of photons in each bin in the latter, while the colored dot-dashed lines show the actual fraction in each of the $5$ frequency bins after re-scaling.  Redder (bluer) lines indicate lower (higher) photon energies, as the annotations indicate.  The re-scaling removes more high-energy photons from $\dot{N}_{\rm em}^{c}$ due to IGM filtering.  {\bf Right}: effective spectral index $\alpha_{\rm eff}$ (see text) for the multi-frequency \textsc{reduced-$c$} and \textsc{full-$c$, re-scaled $\dot{N}_{\rm em}$} simulations.  The latter has a softer spectrum (larger $\alpha_{\rm eff}$) which gets softer with time, reaching $2.5$ by $z = 5$.  }
    \label{fig:spectrum_plot}
\end{figure}

As mentioned, applying Eq.~\ref{eq:main_result_nu} in each frequency bin results in $\dot{N}_{\rm em}^{c}(\nu)$ having a different spectral shape than $\dot{N}_{\rm em}^{\tilde{c}}(\nu)$, in a way that depends on redshift.  We illustrate this in Figure~\ref{fig:spectrum_plot}, where we quantify the spectral shape of $\dot{N}_{\rm em}$ before and after the re-scaling.  The black solid curve in the left panel shows the fraction of the total emissivity in each bin in our multi-frequency \textsc{reduced-$c$} run with $\tilde{c} = 0.2$ (recall that all bins share the same fraction of the photons in this case).  In the right panel, the black dotted curve shows $\frac{1}{N_{\rm freq}}\dot{N}_{\rm em}^{c}$, the fraction that would be in each bin if $\dot{N}_{\rm em}^{c}$ and $\dot{N}_{\rm em}^{\tilde{c}}$ shared the same frequency dependence.  The colored dot-dashed lines show $\dot{N}_{\rm em}^{c}(\nu)$ for each bin, with redder (bluer) colors denoting lower (higher) photon energies (see annotations).  
The re-scaling removes significantly more photons at higher energies, causing $\dot{N}_{\rm em}^{c}(\nu)$ to get softer with decreasing redshift.  

In the right panel, we calculate the ``effective'' power law index, $\alpha_{\rm eff}$, for each case vs. redshift.  We define $\alpha_{\rm eff}$ such that a power law spectrum with that index would have the same $\overline{\sigma_{\rm HI}}$ as the source spectrum in the simulation.  For the \textsc{reduced-$c$} run, this is trivially $1.5$, but for the \textsc{full-$c$, re-scaled $\dot{N}_{\rm em}$} run, $\alpha_{\rm eff}$ increases with time, reaching $\approx 2.5$ by $z = 5$.  Note that this effect arises from the fact that the IGM absorbs (filters) low-energy photons more readily than high-energy ones, such that $\dot{N}_{\gamma}$ has a harder spectrum than $\dot{N}_{\rm em}$.  This effect will thus be less significant in simulations with less filtering.  We showed in Appendix B of Ref.~\cite{Cain2023} that our simulations are likely predict levels of filtering on the high end of expectations, so the effect may be somewhat exaggerated here.  Still, it represents a complication to applying Eq.~\ref{eq:main_result_nu} to multi-frequency simulations that should be carefully considered and quantified in any studies that use this approach.

\section{Conclusions}
\label{sec:conc}

In this work, we have studied the effect of the reduced speed of light approximation (RSLA) on radiative transfer simulations of reionization.  We used a simple analytic model to show (1) when and why the RSLA produces inaccurate results, especially near reionization's end, and (2) that using the RSLA is, to a good approximation, equivalent to a redshift-dependent re-scaling of the global ionizing emissivity of reionization's sources (Eq.~\ref{eq:main_result}).  We have run simulations with the reduced speed of light (\textsc{reduced-$c$} runs) and with the full speed of light after applying the re-scaling prescribed by Eq.~\ref{eq:main_result} (\textsc{full-$c$, re-scaled $\dot{N}_{\rm em}$} runs).  We have assessed how accurate Eq.~\ref{eq:main_result} is by comparing these two sets of simulations in a number of different physical properties and observables.  Our main findings are summarized below: 

\begin{itemize}

    \item For $\tilde{c}$ as small as $0.05$, our \textsc{reduced-$c$} and \textsc{full-$c$, re-scaled $\dot{N}_{\rm em}$} simulations agree in the global reionization history within a linear difference of $\lesssim 0.01$ at all redshifts.  This confirms that the re-scaling of $\dot{N}_{\rm em}$ prescribed by Eq.~\ref{eq:main_result} is equivalent to using the RSLA with respect to the global reionization history.  
    
    \item We have compared the morphology of ionized regions and the 21 cm power spectrum in our \textsc{reduced-$c$} and \textsc{full-$c$, re-scaled $\dot{N}_{\rm em}$} runs.  We find that the sizes of the largest ionized regions are suppressed in the \textsc{reduced-$c$} runs relative to their \textsc{full-$c$, re-scaled $\dot{N}_{\rm em}$} counterparts.  This is caused by position-dependent time-delay effects caused by the RSLA for which Eq.~\ref{eq:main_result} does not account.  The net effect is to suppress ionization power in the \textsc{reduced-$c$} simulations.  We find that differences in the 21 cm power spectrum at $k = 0.1$ $h$Mpc$^{-1}$ is at most $15\%$ ($10\%$) for $\tilde{c} = 0.2$ ($0.3$), but can exceed $20\%$ for $\tilde{c} = 0.1$ and $30\%$ for $\tilde{c} = 0.05$.  As such, we recommend applying Eq.~\ref{eq:main_result} only for $\tilde{c} \geq 0.2$ when ionization morphology is important.  

    \item We have made the same comparison with respect to observables derived from high-redshift quasar spectra, with a focus on the Ly$\alpha$ forest at $z \leq 6$.  We find that when $\dot{N}_{\rm em}$ is calibrated to reproduce the mean transmission of the Ly$\alpha$ forest at $z \leq 6$, the resulting $\dot{N}_{\rm em}$ histories are very different for different $\tilde{c}$.  For \textsc{reduced-$c$} runs calibrated in this way with $\tilde{c} = 0.2$, and $0.3$, we find that their \textsc{full-$c$, re-scaled $\dot{N}_{\rm em}$} agree in the mean transmission to $20\%$ or better.  However, for $\tilde{c} = 0.1$ the difference can be as large as a factor of $2$.  We find qualitatively similar results for the ionizing photon mean free path and the IGM temperature at mean density.  Both agree to within $\leq 5\%$ for $\tilde{c} = 0.2$ and $0.3$, but differ by up to $35\%$ and $6\%$, respectively, for $\tilde{c} = 0.1$.  

    \item We looked at the large-scale fluctuations in the Ly$\alpha$ forest opacity in our two sets of simulations.  For $\tilde{c} \geq 0.2$, the distribution of effective optical depths over $50$ $h^{-1}$Mpc segments of the forest agrees to within a few percent between \textsc{reduced-$c$} and \textsc{full-$c$, re-scaled $\dot{N}_{\rm em}$} simulations.  Once again, we find that these differences are considerably larger for $\tilde{c} = 0.1$.  We explored this result in more depth by looking at the large-scale fluctuations in $\Gamma_{\rm HI}$ and $T$ during the end stages of reionization.  \textsc{reduced-$c$} simulations over-produce large-scale fluctuations in $\Gamma_{\rm HI}$ relative to their \textsc{full-$c$, re-scaled $\dot{N}_{\rm em}$} counterparts thanks to the same position-dependent time-delay effects responsible for affecting the ionization morphology.  In addition, \textsc{reduced-$c$} simulations slightly under-estimate the temperatures of recently ionized, hot voids near reionization's end, owing to their lower mean I-front speeds.  

    \item Lastly, we generalized our method to apply to multi-frequency RT simulations.  We found that with respect to the Ly$\alpha$ forest, the method works just as well with multi-frequency simulations as with single-frequency ones.  However, because of IGM filtering effects, $\dot{N}_{\rm em}^{c}$ ends up having a softer spectrum than $\dot{N}_{\rm em}^{\tilde{c}}$, complicating the physical interpretation of the re-scaling procedure.  We caution that when applying our method to multi-frequency simulations, this effect should be quantified and its implications carefully considered.  
     
\end{itemize}

The approach described in this work is potentially useful in the following situations: either (1) when $\dot{N}_{\rm em}$ is entirely a free function of redshift that is calibrated to match some observable (as in Refs.~\cite{Cain2023,Asthana2024}) or (2) when $\dot{N}_{\rm em}$ is uniquely determined by some set of free parameters that are being marginalized over, as in Ref.~\cite{Qin2021}.  In the first case, $\dot{N}_{\rm em}$ can be calibrated using the RSLA at a reduced computational cost, and Eq.~\ref{eq:main_result} can be applied to the end result to recover the ``true'' $\dot{N}_{\rm em}$.  In the second case, the re-scaling prescribed by Eq.~\ref{eq:main_result} would be equivalent to changing the mapping between $\dot{N}_{\rm em}$ and the free parameters that determine it in a quantifiable way.  This scenario is particularly useful for the goal of constrain the ionizing properties of reionization's sources by combining EoR observables with many parameterized radiative transfer models.  Unfortunately, our method is not directly applicable for situations where $\dot{N}_{\rm em}$ cannot simply be re-scaled after the simulation has been run.  This is the case if $\dot{N}_{\rm em}$ is a self-consistent prediction a model that has no or very few free parameters, such as the THESAN~\citep{Kannan2022,Garaldi2022}, CoDa~\citep{Ocvirk2016,Ocvirk2018}, or SPHINX~\citep{Rosdahl2018,Katz2021} simulations.  Our results further suggest that for the first two applications, the RSLA should only be used (in conjunction with Eq.~\ref{eq:main_result}/\ref{eq:main_result_nu}) for $\tilde{c} \geq 0.2$, to ensure that the effects of the RSLA on large-scale fluctuations in the ionization and radiation fields are minimized.  This corresponds to a speed-up factor of up to $5$, which will help enable considerably faster searches of the reionization parameter space using RT simulations.   

\acknowledgments

The author acknowledges support from the Beus Center for Cosmic Foundations while this work was ongoing.  He also acknowledges helpful conversations with Anson D'Aloisio, Garett Lopez, and Shikhar Asthana.


\bibliography{references.bib}
\bibliographystyle{JHEP}

\end{document}